\documentclass[reprint,showpacs,prepintnumbers,amsmath,amssymb,aps,twocolumn]{aastex63}
\usepackage{amsmath,amssymb}
\usepackage{graphicx}
\usepackage{color}
\usepackage{comment}
\usepackage{bbold}
\usepackage{soul,color}
\sethlcolor{white}
\usepackage{booktabs}
\usepackage{array}
\usepackage{hyperref}
\hypersetup{
    colorlinks=true,       
    linkcolor=blue,          
    citecolor=blue,        
    filecolor=blue,      
    urlcolor=blue           
}

\def\Al{$^{26}$Al}
\def\Cl{$^{34}$Cl}
\def\Kr{$^{85}$Kr}
\def\Mg{$^{26}$Mg}
\def\Cd{$^{113}$Cd}

\newcommand{\nuc}[2]{$^{#1}$#2}

\begin{document}


\accepted{to The Astrophysical Journal Supplement Series}

\hspace{5.2in} \mbox{LA-UR-20-22828}

\title{Astromers: Nuclear Isomers in Astrophysics}

\author[0000-0002-0637-0753]{G.~Wendell Misch}
\email{wendell@lanl.gov}
\affiliation{Theoretical Division, Los Alamos National Laboratory, Los Alamos, NM, 87545, USA}
\affiliation{Center for Theoretical Astrophysics, Los Alamos National Laboratory, Los Alamos, NM, 87545, USA}
\affiliation{Department of Physics, School of Physics and Astronomy, Shanghai Jiao Tong University, Shanghai 200240, China}
\affiliation{Joint Institute for Nuclear Astrophysics - Center for the Evolution of the Elements, USA}

\author[0000-0003-2550-7388]{Surja K. Ghorui}
\affiliation{Quark Matter Research Center, Institute of Modern Physics, Chinese Academy of Sciences, Lanzhou 730000, China}
\affiliation{Department of Physics, School of Physics and Astronomy, Shanghai Jiao Tong University, Shanghai 200240, China}

\author[0000-0002-6389-2697]{Projjwal Banerjee}
\affiliation{Department of Physics, Indian Institute of Technology Palakkad, Palakkad, Kerala 678557, India}

\author[0000-0002-1411-4135]{Yang Sun}
\affiliation{Department of Physics, School of Physics and Astronomy, Shanghai Jiao Tong University, Shanghai 200240, China}
\affiliation{Institute of Modern Physics, Chinese Academy of Sciences, Lanzhou 730000, China}
\affiliation{China Institute of Atomic Energy, P.O. Box 275(10), Beijing 102413, China}

\author[0000-0002-9950-9688]{Matthew~R. Mumpower}
\affiliation{Theoretical Division, Los Alamos National Laboratory, Los Alamos, NM, 87545, USA}
\affiliation{Center for Theoretical Astrophysics, Los Alamos National Laboratory, Los Alamos, NM, 87545, USA}
\affiliation{Joint Institute for Nuclear Astrophysics - Center for the Evolution of the Elements, USA}

\begin{abstract}
We develop a method to compute thermally-mediated transition rates between the ground state and long-lived isomers in nuclei.  We also establish criteria delimiting a thermalization temperature above which a nucleus may be considered a single species and below which it must be treated as two separate species: a ground state species, and an astrophysical isomer (``astromer'') species.  Below the thermalization temperature, the destruction rates dominate the internal transition rates between the ground state and the isomer.  If the destruction rates also differ greatly from one another, the nuclear levels fall out of or fail to reach thermal equilibrium.  Without thermal equilibrium, there may not be a safe assumption about the distribution of occupation probability among the nuclear levels when computing nuclear reaction rates.  In these conditions, the isomer has astrophysical consequences and should be treated a separate astromer species which evolves separately from the ground state in a nucleosynthesis network. We apply our transition rate methods and perform sensitivity studies on a few well-known astromers.  We also study transitions in several other isomers of likely astrophysical interest.
\end{abstract}

\section{Introduction \label{sec:intro}}

Certain excited nuclear states may live much longer than the typical picosecond or femtosecond lifetimes of other states.  The nuclear structure community uses an informal threshold of 1 nanosecond to distinguish these: an excited state which lives for longer than a nanosecond is considered metastable and called a nuclear isomer. The term {\it isomer} in nuclear physics probably came from Frederick Soddy \citep{Soddy1917}, who applied it to nuclei---in analogy with chemical isomers---to describe long-lived nuclear states; we take this term and modify it for those states which impact astrophysics. The discovery of nuclear isomerism is usually credited to the work of Otto Hahn in 1921 in an experiment with uranium \citep{Hahn1921}.  \cite{Weizsacker1936} pointed out that the combination of a large angular momentum change and low transition energy could lead to a long half-life for electromagnetic decay in nuclei.  \cite{BM1953} discussed specific isomeric transitions of the electric quadrupole type.  Today, it is well known that various nuclear structure effects can lead to isomeric states, the best known being spin traps (excited state has a large difference in spin from lower-lying states), K isomers (orientation of spin relative to axis of symmetry in a deformed nucleus is very different from lower-lying states), and shape isomers (excited state is a very different shape from lower-lying states) \citep{WG1999}.

Astrophysical nucleosynthesis calculations rely on accurate nuclear reaction rate inputs.  Indeed, even a single reaction rate can profoundly influence astrophysical evolution (see e.g. \cite{Kirsebom:2019}).  Researchers have therefore done a tremendous amount of work over the decades to compute the rates of nuclear weak interactions \citep{fuller1982stellar,oda1994rate,langanke2001rate}, neutron capture cross sections \citep{dillmann2010kadonis}, and so on.  Typically, one of two treatments is used to compute nucleosynthesis rates.  Either only the ground state rate is used, or the levels are considered to be in a thermal equilibrium probability distribution.  The presence of an isomeric state can diminish the accuracy of both approaches.  When a nucleus is produced, it decays (usually by $\gamma$-ray emission) to a lower state.  If it is caught in an isomer before it reaches ground, it will not necessarily undergo subsequent reactions at the ground state rate.  Isomers can also cause the probability distribution of nuclear energy levels to fail to reach thermal equilibrium.  The destruction rates (e.g. $\beta$ decay) of long-lived states (the ground state and isomers) of a single nuclear species can be vastly different from one another; the most famous example in astrophysics is {\Al}, which has a ground state $\beta$-decay half-life of 717 kyr, but an isomer with a $\beta$-decay half-life of 6.346 s.  If the destruction rates are also fast relative to the transition rates between the long-lived states, a rapidly destroyed state will become depopulated relative to the thermal equilibrium population since the population will tend to stay trapped in the other state.  Although the existence of isomeric states in nuclei has been known for about a century, there remains much to learn about their influence on the creation of the elements in astrophysical nucleosynthesis; they are expected to have significant impact due to their unique decay properties \citep{aprahamian2005long}.

Not all isomers have an impact on astrophysical nucleosynthesis.  Most isomers transition to lower energy states preferentially over destruction channels (e.g. the 103.00 keV isomer in $^{81}$Se, which transitions to ground 99.949\% of the time \citep{baglin2008nuclear}), and sufficient connection to ground ensures destruction cannot cause deviation from thermal equilibrium.  Furthermore, isomers which \emph{can} have an effect will not make a difference in all environments; an isomer may prevent thermalization at low temperature, but in a hotter environment, thermally driven transitions through intermediate states can enable equilibration. There are also isomers that may be isolated from the ground state in the astrophysical site where they are produced, yet are not populated during production of the nucleus (see e.g. \nuc{182}{Hf} in section \ref{sec:$r$-process} and {\Cd} in section \ref{sec:non-astromers}).

We thus see that some isomers can play an influential role in astrophysical nucleosynthesis, but most do not.  This distinction defines astrophysical isomers, or ``astromers'': they are nuclear isomers which have influence as such in an astrophysical environment of interest.

Astromers by definition do not behave the same as their associated ground states.  Consequently, it behooves nucleosynthesis networks to treat them as separate species that can be destroyed and created by transitions into and out of the ground state.  In this work we develop a framework to rigorously treat astromers and these thermally mediated transitions.  We apply our methods to several nuclei of interest in a number of astrophysical sites, including analyses of the effects of uncertainty for three well-known isomers.  We also provide data tables for use in nucleosynthesis networks.

We develop in section \ref{sec:rate_formalism} a highly precise means of calculating effective transition rates between long-lived nuclear states in hot environments.  In section \ref{sec:calculation}, we apply our methods to several well-known and potential astromers, including {\Al}, {\Cl}, and {\Kr}.  We give some further observations and concluding thoughts in section \ref{sec:conclusions}.  Because astromer $\leftrightarrow$ ground transitions are facilitated by intermediate states, it is helpful to trace which intermediate transitions contribute most to the effective transition rate so that we may efficiently assess the effects of uncertainty; appendix \ref{app:pathfinding} describes how we find the greatest contributors and makes important notes about symmetries and detailed balance.

\section{Transition Rate Formalism}
\label{sec:rate_formalism}

Consider a connected system in which every state is reachable from every other state via some set of transitions.  We divide these states into two classes: endpoint states $E$, which in our application are the ground state and isomers in an atomic nucleus; and intermediate states $I$, which for us are non-isomeric excited states.  This discussion specializes to the case of two endpoint states (a single isomer), but the arguments readily generalize.  This section and section \ref{app:pathfinding} ignore destruction of the states via $\beta$ decay, etc.  Destruction will be introduced later as simply another set of rates that do \emph{not} impact internal transition rates.

\begin{figure}
    \centering
    \includegraphics[width=\columnwidth]{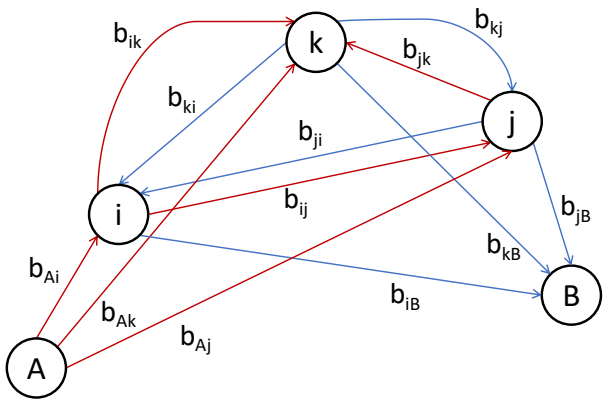}
    \caption{Schematic of transitions from state $A$ to state $B$ when $A$ and $B$ have little or no direct coupling.  If energy is increasing in the upward direction, then red arrows show up transitions and blue arrows show down transitions.  The $b_{st}$ are the probabilities that when the system leaves $s$, it transitions directly to $t$.  Because $A$ cannot transition directly to $B$, transitions between the two require intermediate states.}
    \label{fig:digraph_cartoon}
\end{figure}

We will keep this section general by using the labels $A$ and $B$ for endpoint states and $i$, $j$, etc for intermediate states.  For the sake of concreteness, take $A$ to be the ground state and $B$ to be an isomer of some nucleus.  Figure \ref{fig:digraph_cartoon} provides a schematic for thermally driven transitions from $A$ to $B$ via intermediate states $i$, $j$, and $k$.  When a system transitions out of state $s$, it goes to state $t$ with probability $b_{st}$.

The transition probabilities $b_{st}$ are related to the transition rates $\lambda_{st}$ by
\begin{equation}
    b_{st} = \frac{\lambda_{st}}{\lambda_s \equiv \sum\limits_f \lambda_{sf}}.
    \label{eq:b_definition}
\end{equation}
By inspection, $\sum\limits_t b_{st}=1$, ensuring that there are no sources or sinks in the system.  We compute the $\lambda_{st}$ using the spontaneous $\gamma$-decay rates $\lambda_{hl}^s$ from a higher state $h$ to a lower state $l$.  In this article, we follow the example of \cite{wf:1980} and \cite{coc-etal:1999} and consider only the photon bath.
\begin{align}
    \lambda_{hl} = \lambda_{hl}^s (1+u) \label{eq:lambda_hl} \\
    \lambda_{lh} = \frac{2J_h+1}{2J_l+1}\lambda_{hl}^s u \label{eq:lambda_lh} \\
    u = \frac{1}{e^{(E_h-E_l)/T}-1} \label{eq:u}
\end{align}
These may be combined to obtain the useful relation

\begin{equation}
    \lambda_{st} = \frac{2J_t+1}{2J_s+1} e^{\frac{E_s-E_t}{T}}\lambda_{ts}.
    \label{eq:lambda_det_bal}
\end{equation}
The factor $u$  is the Bose-Einstein distribution of a thermal photon bath and captures the fact that the bath stimulates transitions from $h$ to $l$ and induces transitions from $l$ to $h$.  As \cite{wf:1980} point out, there can be other thermal interactions which affect the $\lambda_{st}$; equations \ref{eq:lambda_hl}-\ref{eq:u} can be extended to include other interactions if needed.  These equations represent our \emph{only} physical assumptions in this section and section \ref{app:pathfinding}.

With the individual transition probabilities, we are now prepared to compute the effective thermal transition rates between endpoint states.

\subsection{Effective Transition Rate}

We express the effective transition rate $\Lambda_{AB}$ from $A$ to $B$ as
\begin{equation}
\Lambda_{AB} = \lambda_{AB} + \sum\limits_i \lambda_{Ai}P_{iB}.
\label{eq:Lam_eff}
\end{equation}
The rate $\lambda_{AB}$ is for transitions directly from $A$ to $B$; this quantity may be negligible, but in some cases it is not, so we include it.  The sum is over intermediate states $i$, and $P_{iB}$ is the probability the system follows a chain of internal transitions from intermediate state $i$ to the endpoint state $B$ without passing through $A$.  We require that the system \emph{not} pass through $A$ as that would be like starting over, amounting to a failed transition.  In effect, equation \ref{eq:Lam_eff} sums the transition rate from $A$ to each other state times that state's probability to go to $B$.

We now must find the $P_{iB}$.  These can be computed by considering the first step of all allowed routes from $i$ to $B$.  A step from $i$ to state $s$ occurs with probability $b_{is}$.  Now, $P_{iB}$ can be identified as the sum over the probabilities of the possible first steps from $i$ folded with the probabilities $P_{sB}$ that the new state eventually reaches $B$.  That is,
\begin{align}
  P_{iB} &= \sum\limits_{s} b_{is}P_{sB},
  \label{eq:prob_comp}
\end{align}

The sum is over all states $s$.  If $s=A$, the transition has failed, and if $s=B$, the transition is complete; therefore, $P_{AB} \equiv 0$ and $P_{BB} \equiv 1$.  States do not transition to themselves, so $b_{ss} \equiv 0 ~\forall ~s$.  These features allow a slight rewrite.
\begin{align}
    P_{iB} = b_{iB} + \sum\limits_j b_{ij}P_{jB}
    \label{eq:P_recursion}
\end{align}
The summation in the second term is now only over intermediate states.  From here, we may express the problem in matrix form.
\begin{align}
  \overrightarrow{P}_{IB} &= \overrightarrow{b}_{IB} + \mathbf{b}_{II}\overrightarrow{P}_{IB} \nonumber \\
  \rightarrow \left( \mathbb{1} - \mathbf{b}_{II} \right)\overrightarrow{P}_{IB} &= \overrightarrow{b}_{IB}
  \label{eq:prob_vec}
\end{align}
In this expression, the subscript $I$ reminds us that the vector and matrix indices run over the intermediate states.  So $\overrightarrow{P}_{IB}$ is the vector with components $P_{iB}$, $\overrightarrow{b}_{IB}$ is the vector with components $b_{iB}$, and $\mathbf{b}_{II}$ is the matrix with elements $b_{ij}$; in each of these, $i$ and $j$ are intermediate states.  Equations \ref{eq:prob_comp} and \ref{eq:prob_vec} represent a system of $N$ linear equations in $N$ variables, where $N$ is the number of intermediate states included in the calculation.  Thus, $\overrightarrow{P}_{IB}$ can be found with a linear equation solver, and inserting $\overrightarrow{P}_{IB}$ into equation \ref{eq:Lam_eff} yields the effective transition rate from $A$ to $B$.

In the case of a system with more than two endpoint states, equation \ref{eq:Lam_eff} holds for all pairs of initial and final endpoint states.  The method is straightforwardly extended by redefining $P_{iE}$ as the probability to eventually reach endpoint state $E$ from intermediate state $i$ without passing through any other endpoints, and all of the same arguments apply.

\subsection{Proof of Solvability}

By the Invertible Matrix Theorem, if $\mathbb{1}-\mathbf{b}_{II}$ is invertible, then a unique solution to equation \ref{eq:prob_vec} exists.
\begin{equation}
  \overrightarrow{P}_{IE} = \left( \mathbb{1} - \mathbf{b}_{II} \right)^{-1} \overrightarrow{b}_{IE}
\end{equation}
To show that that $\mathbb{1}-\mathbf{b}_{II}$ is invertible, we will use the logic of \cite{gupta-meyer:2001}.

We note that the full transition matrix $\mathbf{b}$ with elements $b_{st}$ (which includes endpoint and intermediate states) is a stochastic matrix: its rows sum to unity.  Obviously, $\mathbf{b}$ is non-negative (all entries are real and positive or zero).  Furthermore, because every state is reachable via some path from every other state, $\mathbf{b}$ is irreducible, meaning the rows and columns cannot be permuted to produce a block diagonal form.  If it were reducible, there would be at least two sets of states that communicate only with members of their set, and the graph of the system would not be connected.

Stochastic matrices have an eigenvector with eigenvalue 1; to see this, observe that $\mathbf{b}\overrightarrow{1}=\overrightarrow{1}$, where $\overrightarrow{1}$ is the column vector with all entries equal to 1.  From the Gershgorin Circle Theorem on upper bounds of spectral radii, the largest absolute value among eigenvalues (the spectral radius) of $\mathbf{b}$ is 1.  Because $\mathbf{b}$ is non-negative and irreducible, the Perron-Frobenius Theorem guarantees that there is exactly one eigenvector corresponding to an eigenvalue equal to the spectral radius, {\it i.e.} 1.

We now rely on a corollary to the Perron-Frobenius Theorem: any principle submatrix of an irreducible non-negative matrix has a spectral radius which is strictly less than the full matrix.  A principle submatrix is obtained by deleting rows and columns from a matrix, with the row and column indices being equal.  The matrix $\mathbf{b}_{II}$ is a principle submatrix of $\mathbf{b}$ obtained by deleting the rows containing transitions from endpoints and the columns containing transitions to endpoints.  Therefore, it has a spectral radius less than 1, and all of its eigenvalues have absolute value (modulus) less than 1.

Finally, let $\mathbf{M}$ and $\mathbf{N}$ be square matrices with $\mathbf{N} = \mathbf{M} + \mathbb{1}$.  For every eigenvalue ${^M}\lambda_i$ of $\mathbf{M}$, there will be an eigenvalue ${^N}\lambda_i = {^M}\lambda_i + 1$ of $\mathbf{N}$.  To see this, let $\overrightarrow{x}$ be an eigenvector of $\mathbf{M}$ with eigenvalue $\lambda$.  We then have
\begin{align}
    \mathbf{N}\overrightarrow{x} &= \left( \mathbf{M} + \mathbb{1} \right) \overrightarrow{x} \nonumber \\
    &= \mathbf{M}\overrightarrow{x} + \overrightarrow{x} \nonumber \\
    &= \lambda\overrightarrow{x} + \overrightarrow{x} \nonumber \\
    &= (\lambda + 1) \overrightarrow{x}.
\end{align}
Taking $\mathbf{M}=-\mathbf{b}_{II}$ and remembering that all eigenvalues of $\mathbf{b}_{II}$ have modulus less than 1, we deduce that that all eigenvalues of $\mathbb{1}-\mathbf{b}_{II}$ have a positive real component, are therefore nonzero, and the matrix is invertible by the Invertible Matrix Theorem.

\section{Applications \label{sec:calculation}}

We will now compute the isomer-to-ground and ground-to-isomer effective transition rates and $\beta$-decay rates in several nuclei; all calculations assume an electron density of $\rho Y_e=10^5$ g/cm$^3$.  We will use the prescription from \cite{gupta-meyer:2001} to compute effective decay rates for ensembles of states corresponding to the ground state and the isomer.  For the reader's convenience, we outline the derivation and result here.

Unlike our derivations in section \ref{sec:rate_formalism}, the \cite{gupta-meyer:2001} method assumes intermediate states instantaneously reach steady state equilibrium with the endpoint states (under most circumstances, a reasonable assumption that is backed up by a detailed analysis in that work).  They obtain the steady state occupation fractions $Y$ by noting that the ratio of the intermediate state $i$ and the endpoint state $E$ occupations is related to the ratio of the direct transition rates between them; they call this the ``reverse ratio'' $R_{Ei}$.

\begin{equation}
    R_{Ei} = \frac{\lambda_{Ei}}{\lambda_{iE}} = \frac{Y_i}{Y_E}
    \label{eq:rev_ratio}
\end{equation}

In principle, we would compute this ratio using equation \ref{eq:lambda_det_bal}.  It is undefined if the direct transition rates are zero, but because the ratio is independent of the size of the rates, it remains constant in the limit that the rates approach zero; we find that
\begin{equation}
    R_{Ei} = \frac{2J_i+1}{2J_E+1} e^{\frac{E_E-E_i}{T}}.
\end{equation}

Finally, each intermediate state has the probability $P_{iE}$ to eventually transition to endpoint state $E$ before going to some other endpoint.  \cite{gupta-meyer:2001} interpret this as the fraction of intermediate state $i$ that is associated with endpoint state $E$.  This interpretation---along with the assumption of instantaneous equilibration of intermediate states---enables them to assign a weight $w_{iE}$ to each state and use those weights to construct an ensemble associated with each endpoint.  In effect, every state is assigned a modified Boltzmann factor for each ensemble.
\begin{align}
    w_{iE} &= P_{iE}R_{Ei} = P_{iE} \frac{2J_i+1}{2J_E+1} e^{\frac{E_E-E_i}{T}} \nonumber \\
    w_{E'E} &= \delta_{E'E}
\end{align}

The delta function applies to pairs of endpoint states and ensures that each endpoint contributes only to its own ensemble.  With these weights, we may now compute the effective $\beta$-decay rate $\Lambda_{E \beta}$ (or any other rate) for the ensemble associated with endpoint $E$.
\begin{align}
    \Lambda_{E\beta} = \frac{\sum\limits_s w_{sE}\lambda_{s\beta}}{\sum\limits_s w_{sE}}
    \label{eq:L_EB}
\end{align}
In this expression, $\lambda_{s\beta}$ is the individual $\beta$-decay rate of state $s$.  We are now prepared to compute the transition rates and $\beta$-decay rates of nuclei with isomers.

\subsection{\Al}
\label{sec:calc_al}

{\Al} is produced by massive stars in substantial quantities.  Its ground state $\beta$-decay lifetime of $\sim 1$ million years means that it lives long enough to be observed after a massive star dies, but not so long as to become uncorrelated with where that star died.  Since massive stars have short lifetimes, we infer that if one died recently in a location, that location is probably a region of active star formation.  Hence, {\Al} is an important tracer of star formation \citep{1982ApJ...262..742M,1995A&A...298..445D}, and we are interested in accurately computing its abundance in nucleosynthesis networks.  {\Al} production is sensitive not only to the rates in which it is directly involved, but also to many other reaction rates in nearby nuclei \cite{iliadis-etal:2011}.  However, its treatment is complicated by a long-lived isomer at 228.305 keV that can greatly impact {\Al} destruction rates \citep{ward1980thermalization,coc-etal:1999,gupta-meyer:2001,Runkle_2001,reifarth2018treatment,banerjee-etal:2018}.

To compute {\Al} transition and $\beta$-decay rates, we included 67 total states: the two endpoint states and 65 intermediate states.  In this nucleus and all others in this section, we used available experimental data for the spontaneous transition and $\beta$-decay rates \citep{basunia2016nuclear}.  In {\Al}, we supplemented the data with transition and $\beta$-decay strengths computed using the shell model codes NuShellX \citep{brown2014shell} and OXBASH \citep{oxbash} along with the usdb interaction \citep{br:2006}.  When a transition rate is unmeasured and a shell model calculation is impractical, we use the Weisskopf approximation \citep{weisskopf1930calculation}.  For the $\beta$-decay rates, we used the measured $ft$ values for ground state and isomer $\beta$ decays, and we computed matrix elements with the shell model for all other weak transitions.

\begin{figure}
\includegraphics[width=\columnwidth]{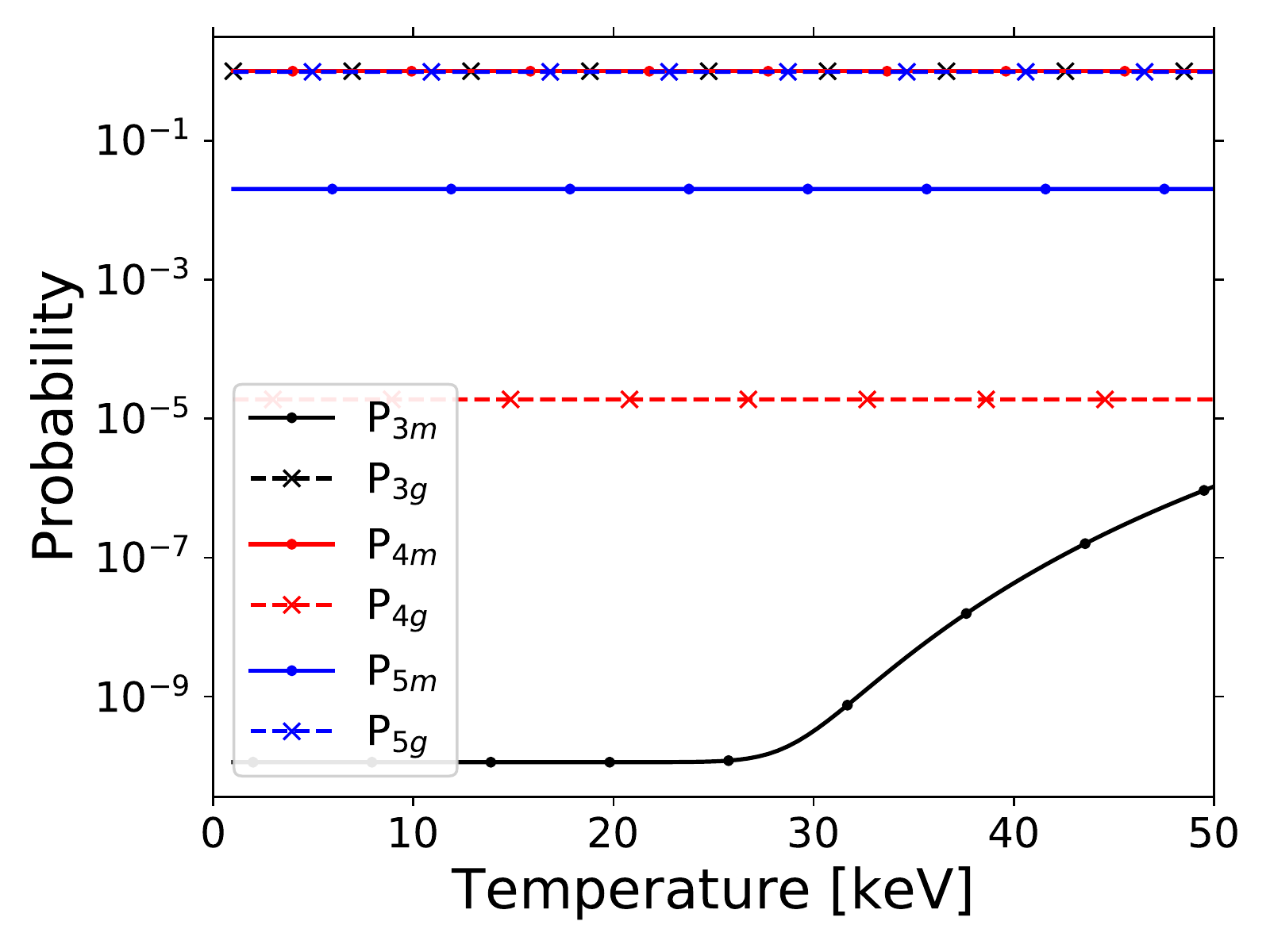}
\caption{The probabilities $P_{iE}$ that intermediate state $i$ reaches endpoint state $E$ before reaching the other endpoint state for the lowest three intermediate states in {\Al}.  Solid (dashed) lines marked with dots (exes) show probabilities to reach the isomer $m$ (ground $g$).  Most are insensitive functions of temperature, though $P_{3m}$ climbs suddenly at $T\approx 30$ keV (see text).}
\label{fig:al_P}
\end{figure}

\begin{figure}
\includegraphics[width=\columnwidth]{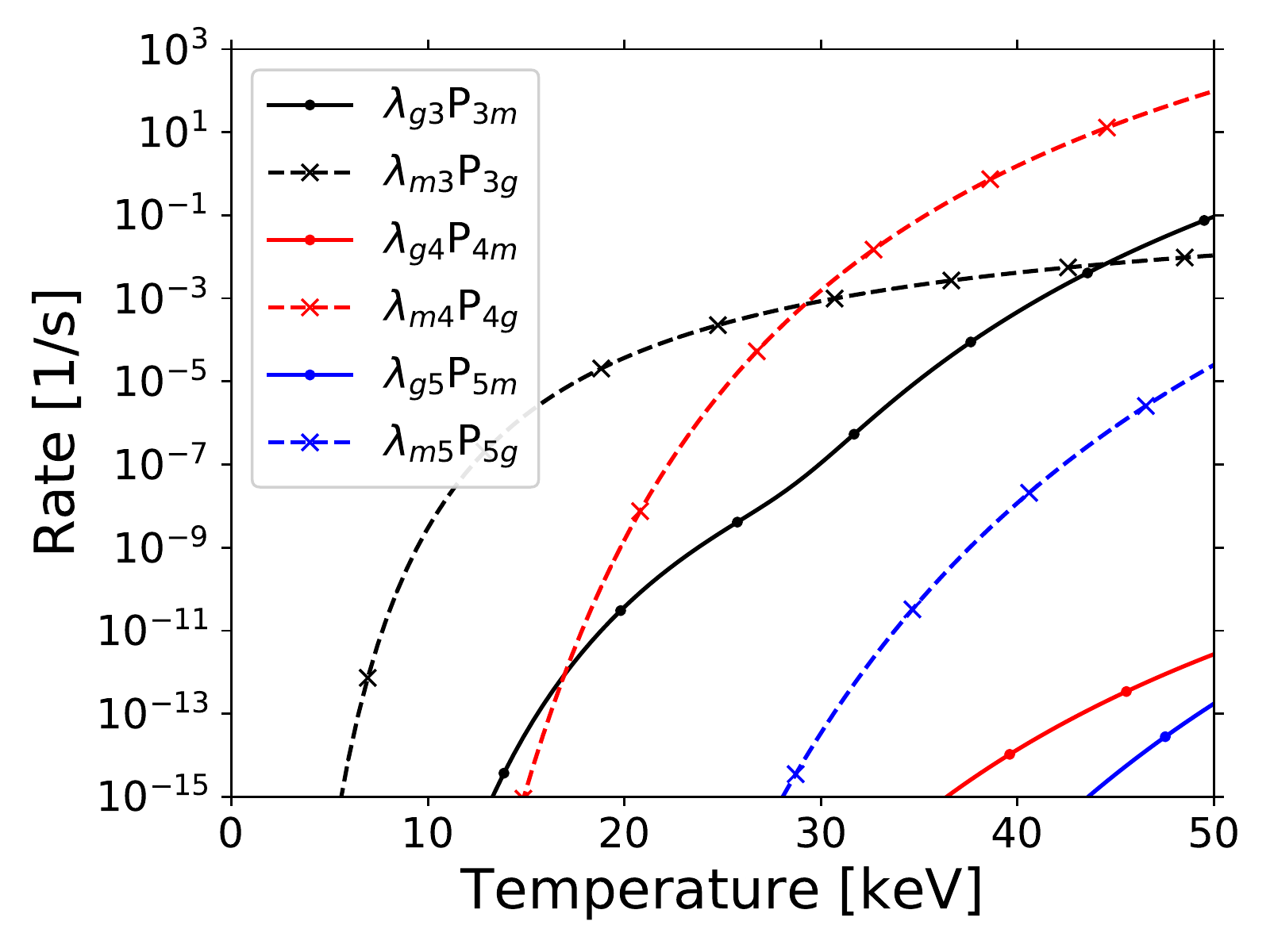}
\caption{Contributions to $\Lambda_{gm}$ and $\Lambda_{mg}$ in {\Al} from individual first steps out of the ground $g$ (solid with dots) and isomeric $m$ (dashed with exes) states.  The contribution is the product of the rate to transition directly from the initial state to the intermediate state and the probability to go from the intermediate state to the destination endpoint state.}
\label{fig:al_problam}
\end{figure}

Once we have the $b_{st}$, obtained from equations \ref{eq:b_definition}-\ref{eq:lambda_det_bal}, we use equation \ref{eq:prob_vec} to compute the $P_{IE}$.  Figure \ref{fig:al_P} shows $P_{IE}$ for the first three intermediate states as functions of temperature; in this figure and figure \ref{fig:al_problam}, the index $g$ indicates the ground state, and $m$ is the isomer.

While most of the $P_{IE}$ are nearly constant, the jump in $P_{3m}$ at around 30 keV portends a behavioral change in the nucleus at this temperature.  Indeed, this manifests as an increase in the rate for the ground state to transition to state 3 and from there ultimately to the isomer.  Figure \ref{fig:al_problam} illustrates this jump; it shows rates for endpoint states to transition to specific intermediate states, and from there to eventually transition to the other endpoint state.  That is, figure \ref{fig:al_problam} shows the terms in equation \ref{eq:Lam_eff}.  Whereas most intermediate states' contributions increase smoothly with temperature, the third state going from $g$ to $m$ (black solid) has a kink at $\sim 30$ keV.

It is no mere coincidence that the kink occurs at the same temperature where $\lambda_{m3}P_{3g}$ (black dashed) crosses $\lambda_{m4}P_{4g}$ (red dashed).  As the temperature rises, state 4 becomes more accessible from the isomer, driving the rise of $\lambda_{m4}P_{4g}$.  Once it dominates $\lambda_{m3}P_{3g}$, the most probable isomer-to-ground transition pathways change\footnote{Transition paths, the algorithm we use to find them, and an important symmetry are detailed in appendix \ref{app:pathfinding}.}.  Figure \ref{fig:al_paths_25_35} shows the three most probable paths from ground to the isomer at $T=25$ keV and 35 keV; by symmetry (section \ref{sec:symmetry}), it also shows the isomer-to-ground paths.

\begin{figure}
    \centering
    \includegraphics[width=\columnwidth]{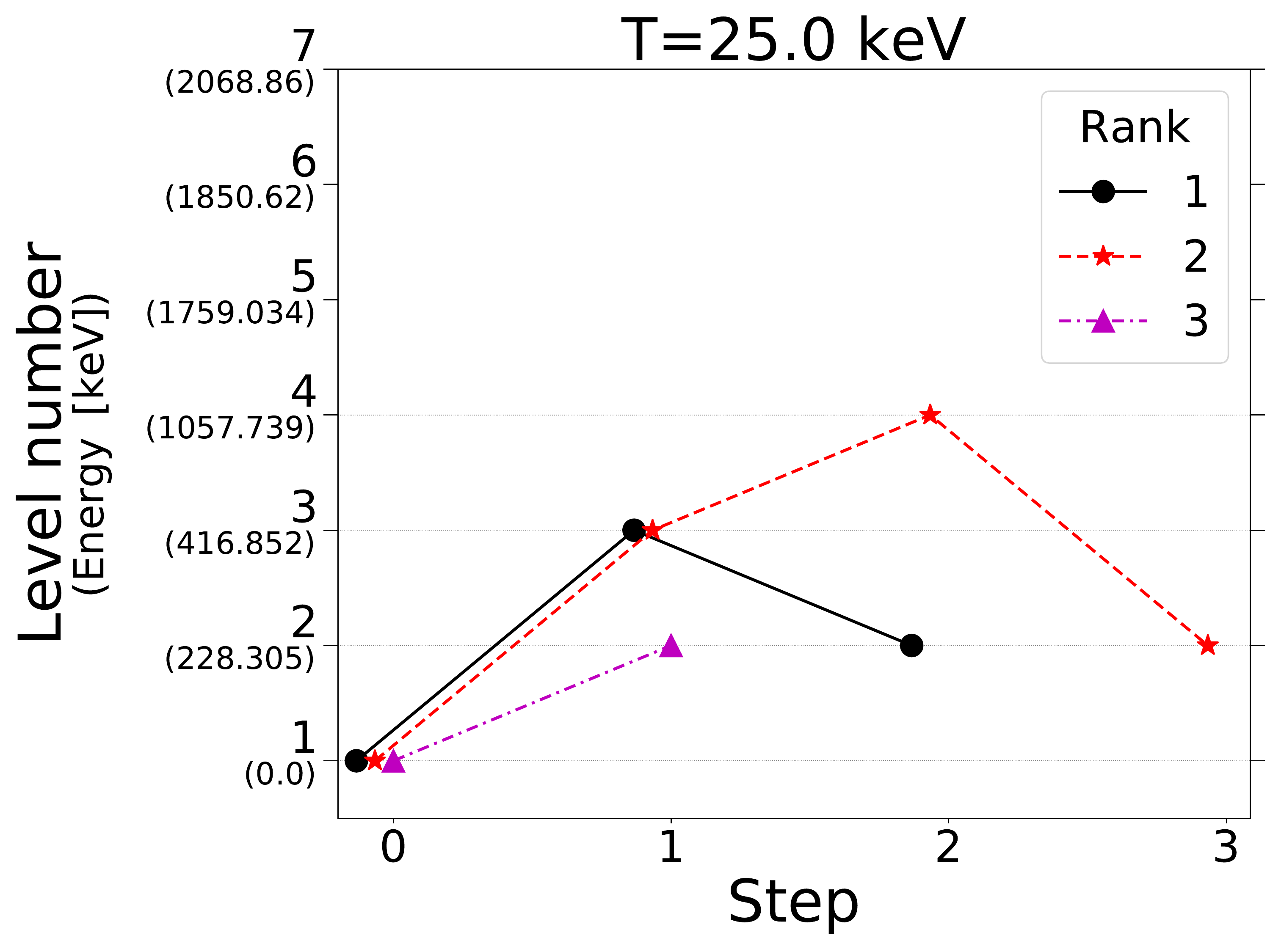}
    \includegraphics[width=\columnwidth]{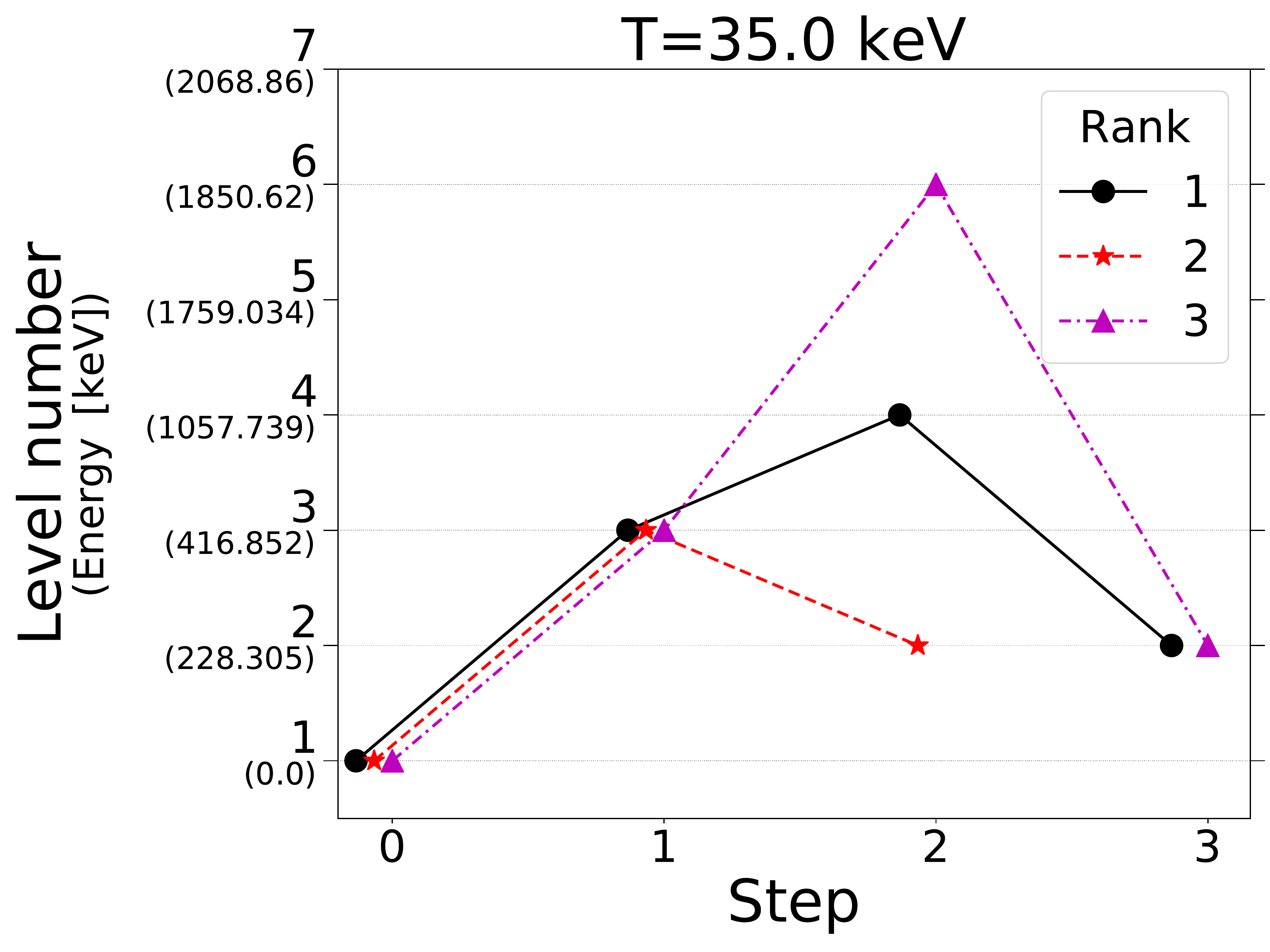}
    \caption{The three paths giving the greatest contribution to the transition rate between the ground state and isomer in {\Al} at temperature $T=25$ keV (top) and $T=35$ keV (bottom).  Although only transitions from 1 to 2 are shown, path reversal symmetry guarantees that the routes from 2 to 1 are the same.}
    \label{fig:al_paths_25_35}
\end{figure}

\begin{figure}
\includegraphics[width=\columnwidth]{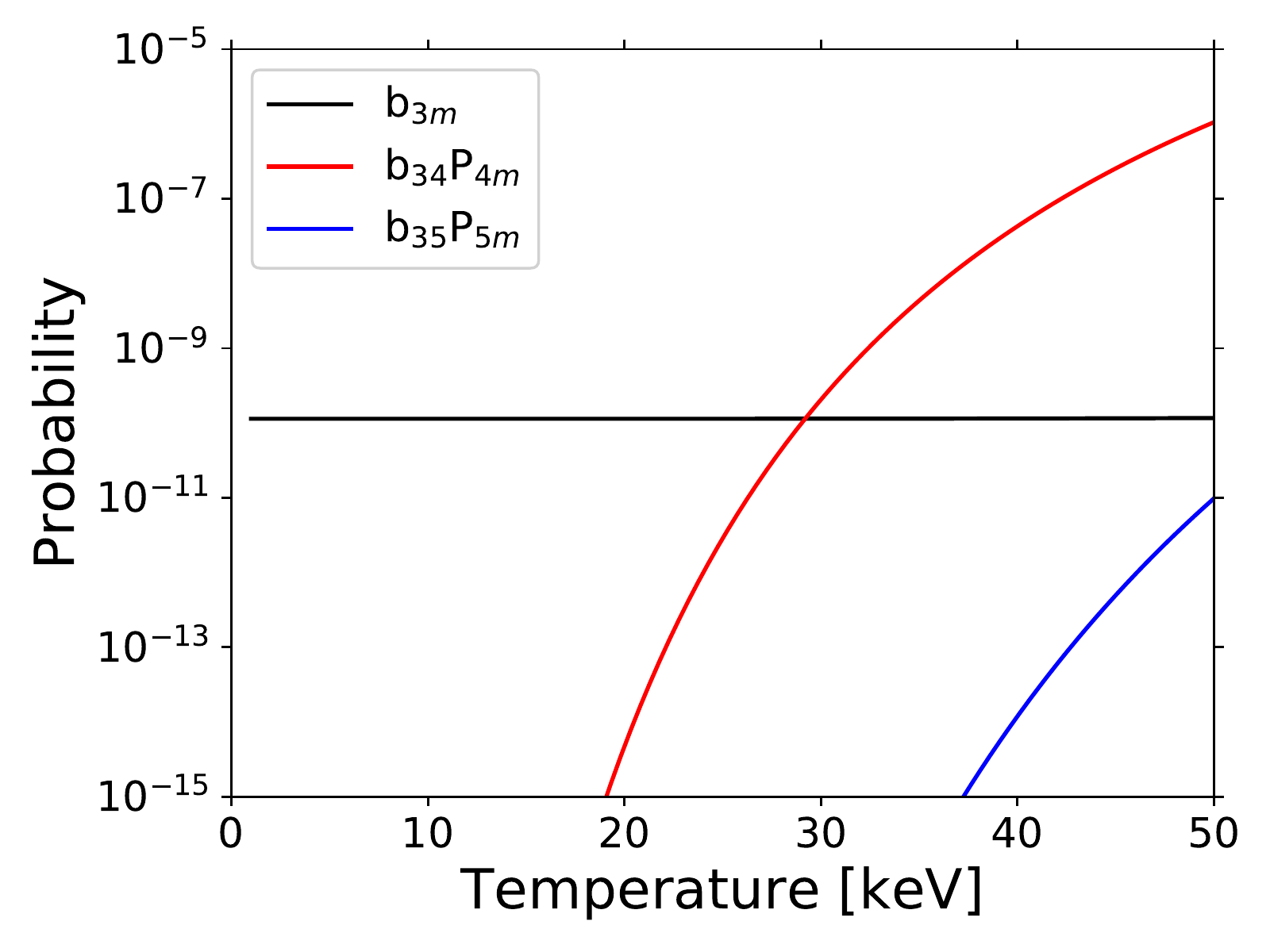}
\caption{Probabilities to follow routes from the third state to the isomer $m$ in {\Al}.  Once the nucleus is in the third state, $b_{3m}$ is the probability that it transitions directly to the isomer, while $b_{3i}P_{im}$ is the probability to go to state $i$ and from there via some path that leads to $m$ without going through ground.}
\label{fig:al_3_prob}
\end{figure}

At $T=25$ keV, the dominant pathway between 1 (ground) and 2 (isomer) is simply through state 3.  But at $T=35$ keV, state 4 is thermally accessible from 2, and because it is much more strongly coupled to the isomer than is state 3 ($P_{4m} >> P_{3m}$ in figure \ref{fig:al_P})\footnote{To be precise, we should use $b_{3m}$, $b_{4m}$, et cetera rather than $P$.  Nevertheless, $P$ is adequate for the present qualitative analysis.}, 2 preferentially transitions to 4.  State 4 is only weakly coupled to ground ($P_{4g}$ in figure \ref{fig:al_P}), and the preferred route from 4 to 1 is via 3.  Therefore, when the isomer preferentially transitions to state 4 (above $T=30$ keV), the most probable isomer-to-ground path is $2 \rightarrow 4 \rightarrow 3 \rightarrow 1$.  This ``opens up'' the $1 \rightarrow 3 \rightarrow 4 \rightarrow 2$ path and creates the kinks.  Figure \ref{fig:al_3_prob} further illustrates this point by showing the probabilities for state 3 to go directly to the isomer ($b_{3m}$), to go to 4 and then eventually to the isomer ($b_{34}P_{4m}$, and to go to 5 and then eventually to the isomer ($b_{35}P_{5m}$).

\begin{figure}
    \includegraphics[width=\columnwidth]{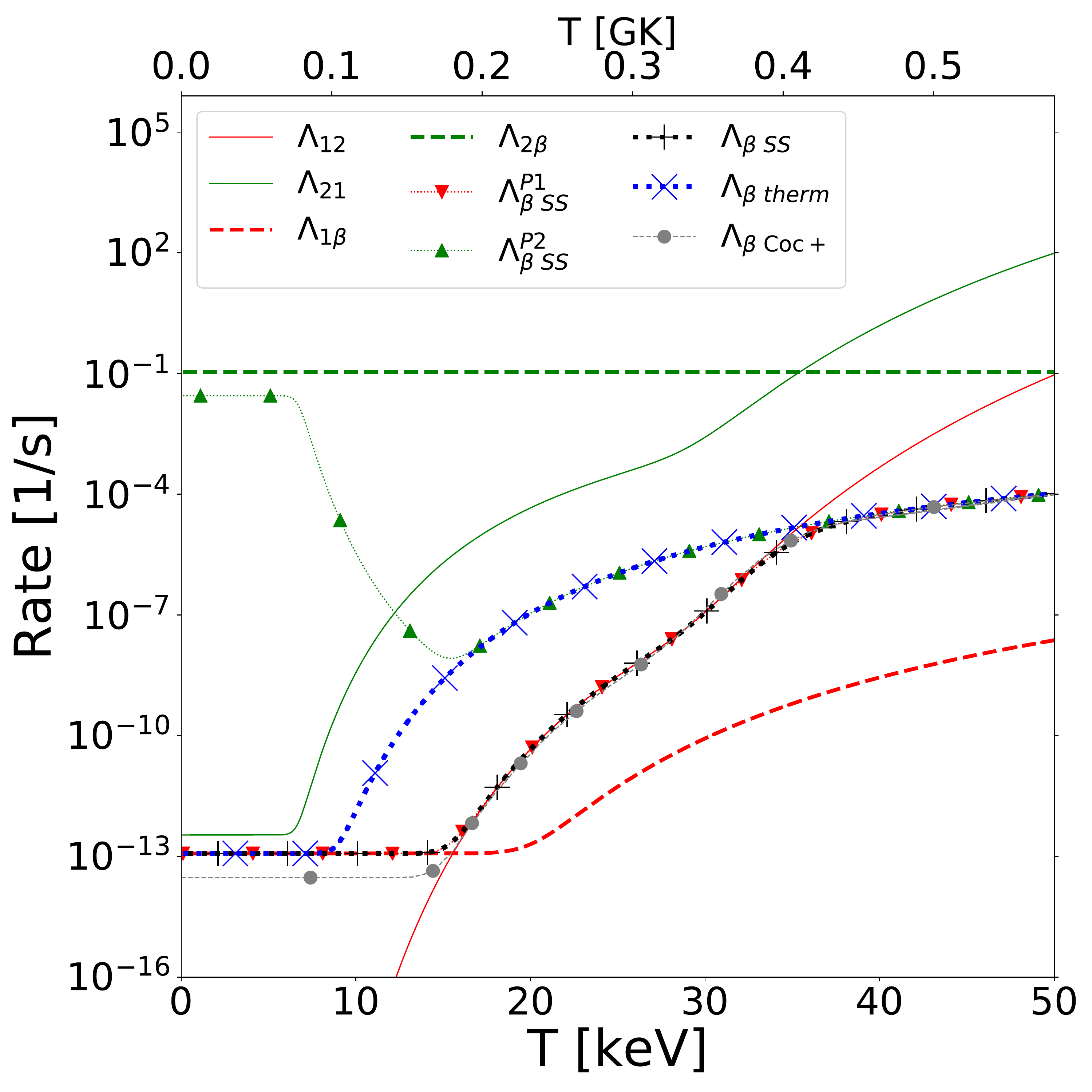}
    \caption{{\Al} transition rates and $\beta$-decay rates.  Solid lines with no markers show the effective ground $\leftrightarrow$ isomer transition rates, and other lines show $\beta$-decay rates.  Red: Ground state ensemble transition and $\beta$-decay rates; down triangles show the steady-state decay rate when only the ground state is produced.  Green: Isomer ensemble transition and $\beta$-decay rates; up triangles show the steady-state decay rate when only the isomer is produced.  Black (crosses): Steady-state $\beta$-decay rate in the absence of production.  Blue (exes): Thermal equilibrium $\beta$-decay rate.  Gray (circles): Effective decay rate from \cite{coc-etal:1999}.}
    \label{fig:al_lambdas}
\end{figure}

Finally, we use equation \ref{eq:Lam_eff} to compute the effective transition rates $\Lambda_{12}$ and $\Lambda_{21}$, shown in figure \ref{fig:al_lambdas}.  From here on, 1 should be understood as the \emph{ensemble} corresponding to the ground state, and 2 is the \emph{ensemble} corresponding to the isomer; the ensembles are constructed as described at the beginning of this section in equations \ref{eq:rev_ratio}-\ref{eq:L_EB}.  The figure also shows various $\beta$-decay rates: the ensemble decay rates $\Lambda_{1\beta}$ and $\Lambda_{2\beta}$; the thermal decay rate $\Lambda_{\beta~therm}$, computed using a thermal distribution of level occupations; the steady-state decay rate $\Lambda_{\beta~SS}$; the steady-state decay rates $\Lambda_{\beta~SS}^{P1}$ and $\Lambda_{\beta~SS}^{P2}$ when the nucleus is \emph{produced exclusively} in either the ground state ($P1$) or isomer ($P2$) ensemble; and---for comparison---the effective decay rate found by \cite{coc-etal:1999}.  We computed the steady-state rates with the formalism of \cite{mgs:2018} assuming a two-level system comprised of 1 and 2.  At low temperatures, we compute a slightly higher $\beta$-decay rate than \cite{coc-etal:1999}; this is because they used laboratory rates, whereas our electron capture rates are enhanced by the dense environment.  We are otherwise in good agreement.

Let us first understand $\Lambda_{12}$, $\Lambda_{21}$, $\Lambda_{1\beta}$, and $\Lambda_{2\beta}$.  Below $T\approx 7$ keV, excited states are inaccessible.  The isomer transitions directly to ground at a slow trickle, the ground state cannot reach the isomer, and each $\beta$ decays at its laboratory rate.  Above $T\approx 7$ keV, state 3 becomes accessible, the path $1 \leftrightarrow 3 \leftrightarrow 2$ opens up, and the transition rates begin to rise.  At $T\approx 13$ keV, $\Lambda_{12}$ surpasses $\Lambda_{1\beta}$, and the dominant ``destruction'' channel for ensemble 1 is transitions to ensemble 2.  At $T\approx 17$ keV, the weight $w_{31}$ of state 3 in ensemble 1 grows great enough that it contributes significantly to the ensemble $\beta$-decay rate.  State 3 has allowed decays to \Mg{}, and $\Lambda_{1\beta}$ rises accordingly.  At $T\approx 30$ keV, the path $1 \leftrightarrow 3 \leftrightarrow 4 \leftrightarrow 2$ opens up, and the transition rates rise faster.  At $T \approx 35$ keV, $\Lambda_{21}$ surpasses $\Lambda_{\beta 2}$, and the dominant destruction channel for ensemble 2 is transitions to ensemble 1.

Now let's examine $\Lambda_{\beta~therm}$ and $\Lambda_{\beta~SS}$, which are the thermal-equilibrium $\beta$-decay rate and the steady-state (without any {\Al} production) $\beta$-decay rate, respectively.  The thermal decay rate is straightforward to understand.  If {\Al} had no isomer, the nuclear level occupation probabilities would simply follow a Boltzmann distribution.  At $T\approx 8$ keV, the first excited state would be sufficiently populated to contribute to $\Lambda_{\beta~therm}$, and the decay rate would rise concomitantly.  However, {\Al} \emph{does} have an isomer which suppresses thermally-driven transitions up from ground.  Because there is no thermally accessible path to populate the isomer, it decays away rapidly, is not replenished, and has a less-than-thermal population in steady state; this suppresses $\Lambda_{\beta~SS}$.  Once $\Lambda_{12}$ exceeds $\Lambda_{1\beta}$ at $T\approx 13$ keV, the ground state begins to appreciably feed the isomer.  But up until $T\approx 35$ keV, the isomer essentially always $\beta$ decays.  Furthermore, $\Lambda_{2\beta} >> \Lambda_{1\beta}$, so {\Al} decays as fast as the isomer is fed, and the steady-state $\beta$-decay rate is approximately equal to $\Lambda_{12}$.  At $T\approx 35$ keV, when $\Lambda_{21}$ exceeds $\Lambda_{2\beta}$, transitions between the ensembles dominate the ensemble decay rates, and the system at last reaches the thermal equilibrium found by other authors \citep{coc-etal:1999,iliadis-etal:2011,banerjee-etal:2018}.  This defines the thermalization temperature below which the {\Al} isomer is an astromer.

This leaves us to understand $\Lambda_{\beta~SS}^{P1}$ and $\Lambda_{\beta~SS}^{P2}$, which are the steady-state decay rates when {\Al} is produced solely in the ground state or isomer ensembles, respectively.  Observe that $\Lambda_{\beta~SS}^{P1}$ tracks $\Lambda_{\beta~SS}$ at all temperatures.  This is because in steady state without production, essentially all of the population is in the ground state anyway, so it makes no difference if {\Al} is produced in the ground state ensemble.

In contrast, when {\Al} is produced in the isomer ensemble at low temperature, the ground state is fed exceedingly slowly.  Gradually, a small equilibrium amount of material will build up in the ground state that brings down the total decay rate, though this may take many years (see \cite{mgs:2018} figure 5), and it will never be enough to bring the decay rate to near the ground state rate.  At $T\approx 7$ keV, $\Lambda_{21}$ increases dramatically, so the ground state is more rapidly fed.  The increased feeding yields a larger quantity of ground state material, and the steady-state decay rate consequently falls.  Once $\Lambda_{12}$ overtakes $\Lambda_{1\beta}$ at $T\approx 13$ keV, the steady-state decay rate goes to the thermal equilibrium rate.  This follows from equation 18 in \cite{mgs:2018} using our two-ensemble system.  That equation is expressed in matrix notation, but we may analyze one of the embedded scalar equations by selecting a single vector component.  Taking the first element of the vectors, we have

\begin{equation}
    \Lambda_{21}n_2 - \Lambda_{12}n_1 - \Lambda_{1\beta}n_1 + \left( \Lambda_{1\beta}n_1 + \Lambda_{2\beta}n_2\right)p_1 = 0.
\end{equation}

Here $n_1$ and $n_2$ are the occupation fractions of the two ensembles, and $p_1$ is the fraction of {\Al} production that goes into the ground state ensemble.  By assumption $p_1=0$, and we are considering a regime where $\Lambda_{12}$ dominates $\Lambda_{1\beta}$; the latter point implies that $\Lambda_{12}n_1 + \Lambda_{1\beta}n_1 \approx \Lambda_{12}n_1$.  Using these facts along with equation \ref{eq:effective_det_bal}, we find

\begin{align}
     & \Lambda_{21}n_2 - \Lambda_{12}n_1 \approx 0 \nonumber \\
     \rightarrow & \frac{n_2}{n_1} \approx \frac{\Lambda_{12}}{\Lambda_{21}} = \frac{g_2}{g_1}e^\frac{E_1-E_2}{T},
\end{align}
which is precisely a thermal distribution between the ground state and isomer.  We conclude that the nuclear levels are thermally distributed when $\Lambda_{12} >> \Lambda_{1\beta}$, leading to the observed agreement between $\Lambda_{\beta~SS}^{P2}$ and $\Lambda_{\beta~therm}$.  Note well that this is \emph{not} necessarily an excuse to just use the thermal decay rate, since it may take many thousands of years to reach steady state.

\begin{figure}
    \centering
    \includegraphics[width=\columnwidth]{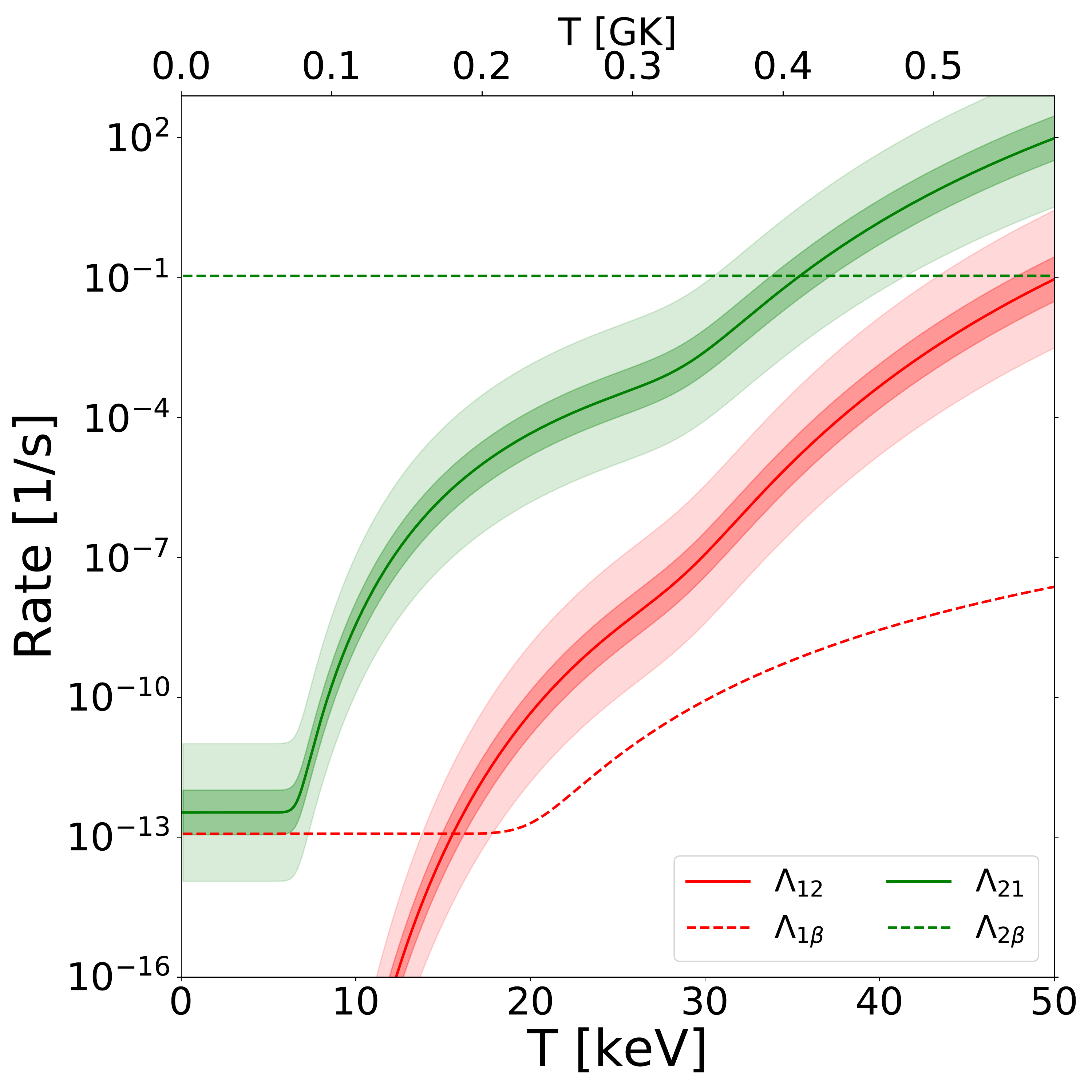}
    \caption{Range of {\Al} effective transition rates.  Dark bands: measured rates increased/decreased by one standard deviation, shell model rates multiplied/divided by a factor of 3.  Light bands: measured rates increased/decreased by two standard deviations, shell model rates multiplied/divided by a factor of 30.}
    \label{fig:26al_uncertainty}
\end{figure}

Up to now, we have used exact values for the $\lambda_{st}$.  However, uncertainties in the individual rates lead to uncertainty in the effective transition rates.  We therefore performed a sensitivity study by varying the $\lambda_{st}$. Figure \ref{fig:26al_uncertainty} shows the results of adjusting measured rates \emph{all} up or \emph{all} down by 1 or 2 standard deviations and shell model rates \emph{all} up or \emph{all} down by a factor of 3 or 30.  Shell model rates using the USDB Hamiltonian tend to be good to within a factor of a couple with occasional large errors relative to experiment of one to two orders of magnitude (see \cite{banerjee-etal:2018}), so this choice of variations should reasonably probe the range of the $\Lambda_{AB}$.  Dark bands show the smaller variations, and light bands show the larger.  Because the $\lambda_{st}$ were adjusted up and down together, the bands represent reasonable bounds on the effective rates.  Furthermore, our analysis finds that the thermalization temperature of {\Al} is uncertain within the range between 30 and 40 keV; this is at odds with \cite{coc-etal:1999}, who concluded that the thermalization temperature is insensitive to remaining nuclear uncertainties.

A more detailed picture emerges when the $\lambda_{st}$ are varied separately.  We identified which individual rates to change by examining the dominant transition paths; if a transition existed in paths carrying a total of at least 1\% of the effective rate and that transition's uncertainty was at least 10\%, we varied it.  We adjusted measured rates up and down by one standard deviation and shell model rates up and down by a factor of 3.  Table \ref{tab:26al_sensitivity} shows the sensitivity of the effective transition rate to variations in the individual rates.

\begin{table}[]
    \centering
    \caption{Most-uncertain individual transitions in the dominant pathways through {\Al}.  The type column indicates whether this spontaneous transition rate comes from shell model calculations (SM) or from experiment (Exp).  The column labeled Fraction shows what fraction of the effective transitions flow through paths containing the individual transition (or its reverse).  Variation is an estimate of the uncertainty; we use the published uncertainty for experimental rates, a factor of 3 for SM rates, and a factor of 10 for Weisskopf rates.  The last column shows the fractional change in effective transition rates when the individual rate is adjusted up and down by one ``unit'' of uncertainty.}
    \begin{tabular}{c|c|c|c|c|c}
        T & Trans & Type & Fraction & Variation & Impact \\
        \hline\hline
        10 & $3 \rightarrow 2$ & SM & 0.9999 & $\times 3$ & 0.3334--2.9997 \\
        \hline
        15 & $3 \rightarrow 2$ & SM & 1.0000 & $\times 3$ & 0.3333--3.0000 \\
        \hline
        20 & $3 \rightarrow 2$ & SM & 1.0000 & $\times 3$ & 0.3334--2.9998 \\
        \hline
        25 & $3 \rightarrow 2$ & SM & 0.9800 & $\times 3$ & 0.3466--2.9602 \\
           & $4 \rightarrow 2$ & Exp & 0.0199 & $20\%$ & 1.0 \\
           & $4 \rightarrow 3$ & SM & 0.0199 & $\times 3$ & 0.9867--1.0399 \\
        \hline
        30 & $3 \rightarrow 2$ & SM & 0.4070 & $\times 3$ & 0.7287--1.8139 \\
           & $4 \rightarrow 2$ & Exp & 0.5930 & $20\%$ & 1.0 \\
           & $4 \rightarrow 3$ & SM & 0.5930 & $\times 3$ & 0.6046--2.1860 \\
        \hline
        35 & $3 \rightarrow 2$ & SM & 0.0315 & $\times 3$ & 0.9790--1.0630 \\
           & $4 \rightarrow 2$ & Exp & 0.9685 & $20\%$ & 1.0 \\
           & $4 \rightarrow 3$ & SM & 0.9685 & $\times 3$ & 0.3543--2.9369 \\
        \hline
        40 & $3 \rightarrow 2$ & SM & 0.0033 & $\times 3$ & 0.9978--1.0066 \\
           & $4 \rightarrow 2$ & Exp & 0.9967 & $20\%$ & 1.0 \\
           & $4 \rightarrow 3$ & SM & 0.9967 & $\times 3$ & 0.3355--2.9933 \\
    \end{tabular}
    \label{tab:26al_sensitivity}
\end{table}

Our results generally agree quite well with \cite{gupta-meyer:2001}.  The greatest uncertainty in the effective transition rates arise from $\lambda_{32}$ at lower temperatures and $\lambda_{43}$ at higher temperatures.

\subsection{\Cl}

{\Cl} may be observable immediately after a nova \citep{coc-etal:1999}, which means that accurately computing its abundance could be interesting.  However, it has a long-lived isomer at 146.36 keV, bringing with it the familiar difficulty.  But unlike {\Al} and the other nuclei in this paper, the {\Cl} isomer is \emph{more} $\beta$-stable than the ground state; the ground state $\beta$-decay half-life is 1.5266 s, while the isomer's total half-life is 31.99 min \citep{ns:2012}.

\begin{figure}
    \centering
    \includegraphics[width=\columnwidth]{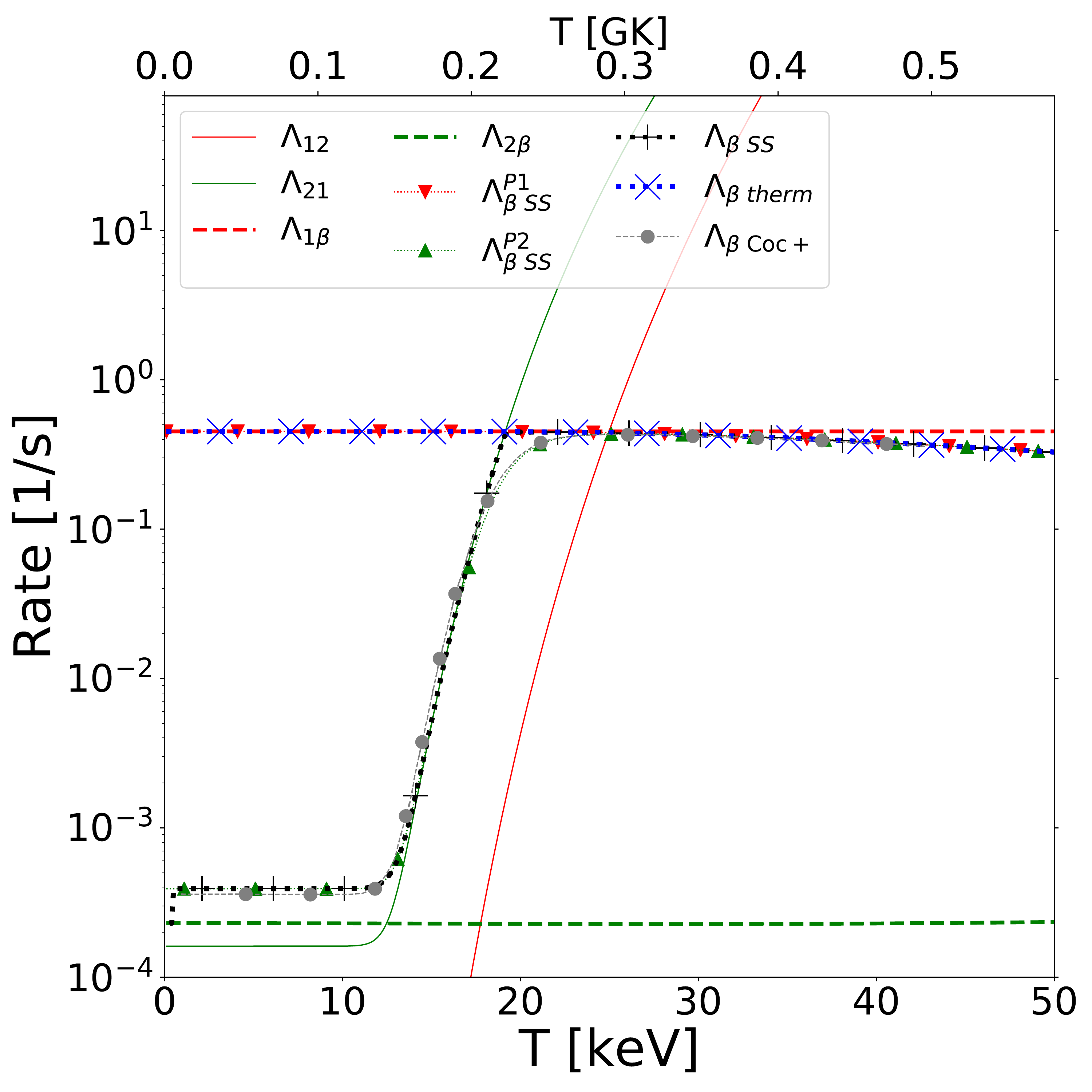}
    \caption{\Cl{} transition and $\beta$-decay rates.  The lines are as in figure \ref{fig:al_lambdas}.}
    \label{fig:cl_rates}
\end{figure}

We computed the {\Cl} rates using 30 states.  As with {\Al}, we supplemented unmeasured transition rates and $\beta$-decay $ft$ values with calculations from NuShellX, OXBASH, and the usdb interaction.  Experimental data for the key $3 \rightarrow 2$ transition consists only of an upper bound, so we used our shell model result.  Our computed effective transition rates and $\beta$-decay rates are shown in figure \ref{fig:cl_rates} along with the rates computed by \cite{coc-etal:1999}.  As with \nuc{26}{Al}, we are in good agreement with the earlier work.

Below $T\approx 10$ keV, no paths apart from the direct transition contribute meaningfully to the transition rate, but above this temperature, the path $2 \leftrightarrow 3 \leftrightarrow 1$ opens up and $\Lambda_{21}$ begins to rise.  Because 2 is the more $\beta$-stable state, $\Lambda_{\beta~SS}^{P2}$ mostly tracks $\Lambda_{\beta~SS}$.  Once $\Lambda_{21}$ dominates $\Lambda_{2\beta}$, $\Lambda_{\beta~SS}$ follows it until joins $\Lambda_{\beta~therm}$, while $\Lambda_{\beta~SS}^{P1}$ is delayed slightly; {\Cl} thermalizes at $T\approx 20$ keV.  Below this temperature, the isomer is an astromer.  With ground being the less stable state, $\Lambda_{\beta~therm}$ and $\Lambda_{\beta~SS}^{P1}$ both track $\Lambda_{1\beta}$ until $T\approx 30$ keV, at which point the isomer ensemble is thermally populated at a sufficient level to pull the thermal and steady state rates down a bit.

\begin{figure}
    \centering
    \includegraphics[width=\columnwidth]{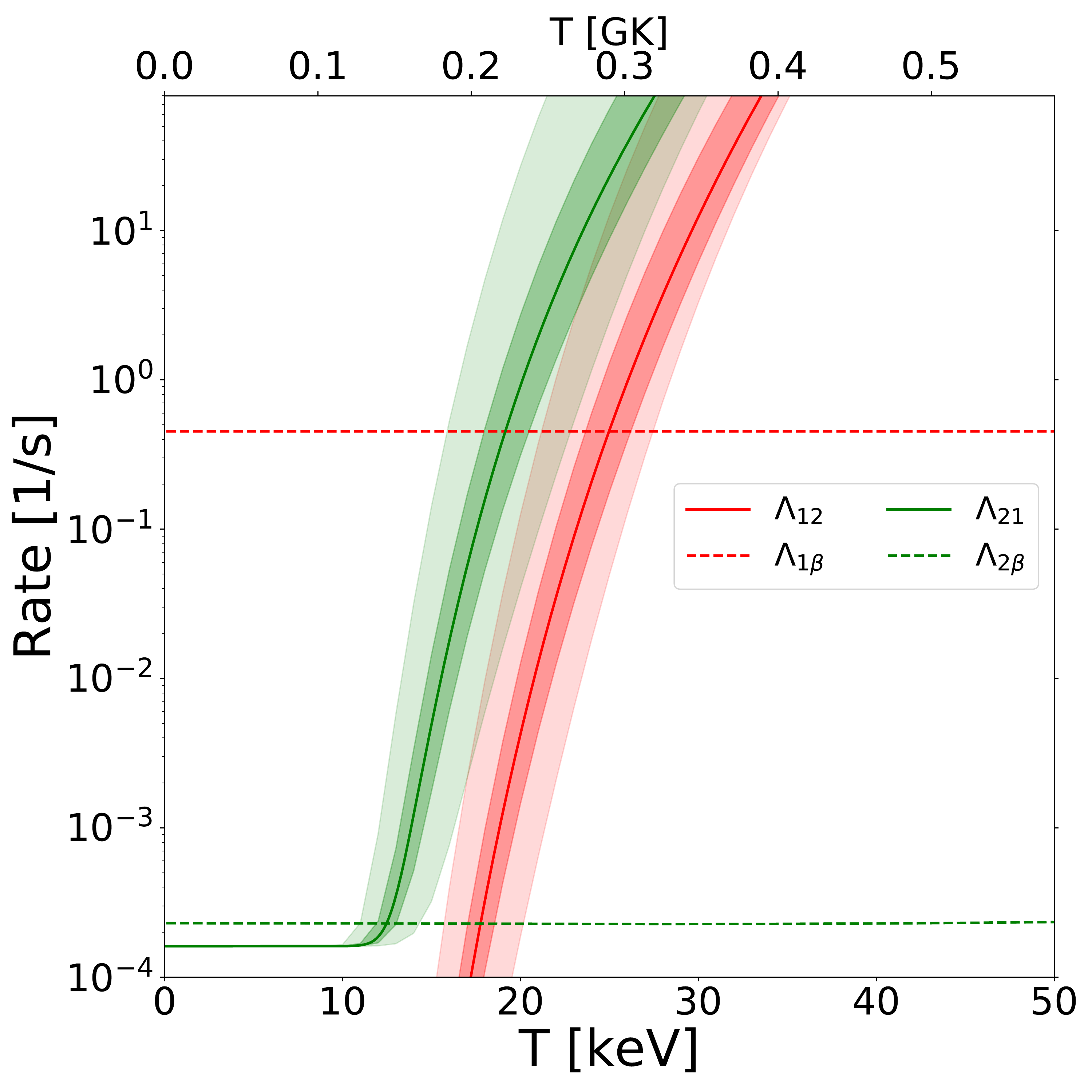}
    \caption{Range of {\Cl} effective transition rates.  Dark bands: measured rates increased/decreased by one standard deviation, shell model rates multiplied/divided by a factor of 3.  Light bands: measured rates increased/decreased by two standard deviations, shell model rates multiplied/divided by a factor of 30.  The uncertainties are dominated by the $3 \rightarrow 2$ transition.}
    \label{fig:34cl_uncertainty}
\end{figure}

Figure \ref{fig:34cl_uncertainty} shows the uncertainty bands in the {\Cl} $\Lambda_{AB}$.  The uncertainty is driven almost entirely by the $3 \rightarrow 2$ transition.  Even up to $T=50$ keV, all dominant paths travel through states no higher than state 4, and apart from $\lambda_{32}$, the $\lambda_{st}$ are measured with small experimental uncertainties.  Our dark uncertainty bands (smaller variations) agree well with the findings of \cite{coc-etal:1999}.  The light bands (larger variations) exhibit some asymmetry: the lower bands are narrower than the upper bands at higher temperatures.  As $\lambda_{32}$ is increased, the nucleus flows more freely through the $1 \rightarrow 3 \rightarrow 2$ path, and $3 \rightarrow 2$ is the bottleneck.  But when $\lambda_{32}$ is sufficiently decreased, the $1 \rightarrow 4 \rightarrow 2$ path dominates at higher temperatures, and the effective transition rate becomes insensitive to further decrease in $\lambda_{32}$.

\subsection{\Kr{}}
\label{sec:kr}

In slow neutron capture ($s$-process) nucleosynthesis, neutron captures are generally slow relative to $\beta$ decays, so few nuclei are produced which are more than one or two steps from stability.  In most cases, this makes it easy to trace out the $s$-process path in the chart of nuclides.  Starting with a seed nucleus, add neutrons until you make a $\beta$-unstable nucleus.  Allow that unstable nucleus to $\beta$ decay until it's stable.  Resume adding neutrons and repeat until you have lead, bismuth, or no more neutron source.

{\Kr} lies along the $s$-process nucleosynthesis path, but the simple procedure above is derailed by the ground state's long-but-not-too-long $\beta$-decay half-life of 10.739 years.  This creates competition between $\beta$ decay and neutron capture which influences the production of nearby nuclei \citep{abia200185kr}.  Because the $s$-process path can go in two different directions at {\Kr}, it is known as a branch-point nucleus and serves as an interesting diagnostic of the $s$-process environment.  However, its current usefulness as such is greatly diminished by two complications: an uncertain neutron capture cross section \citep{dillmann2010kadonis}, and an isomer at 304.871 keV with a total half-life of 4.480 hours \citep{sc:2014}.

\begin{figure}
    \centering
    \includegraphics[width=\columnwidth]{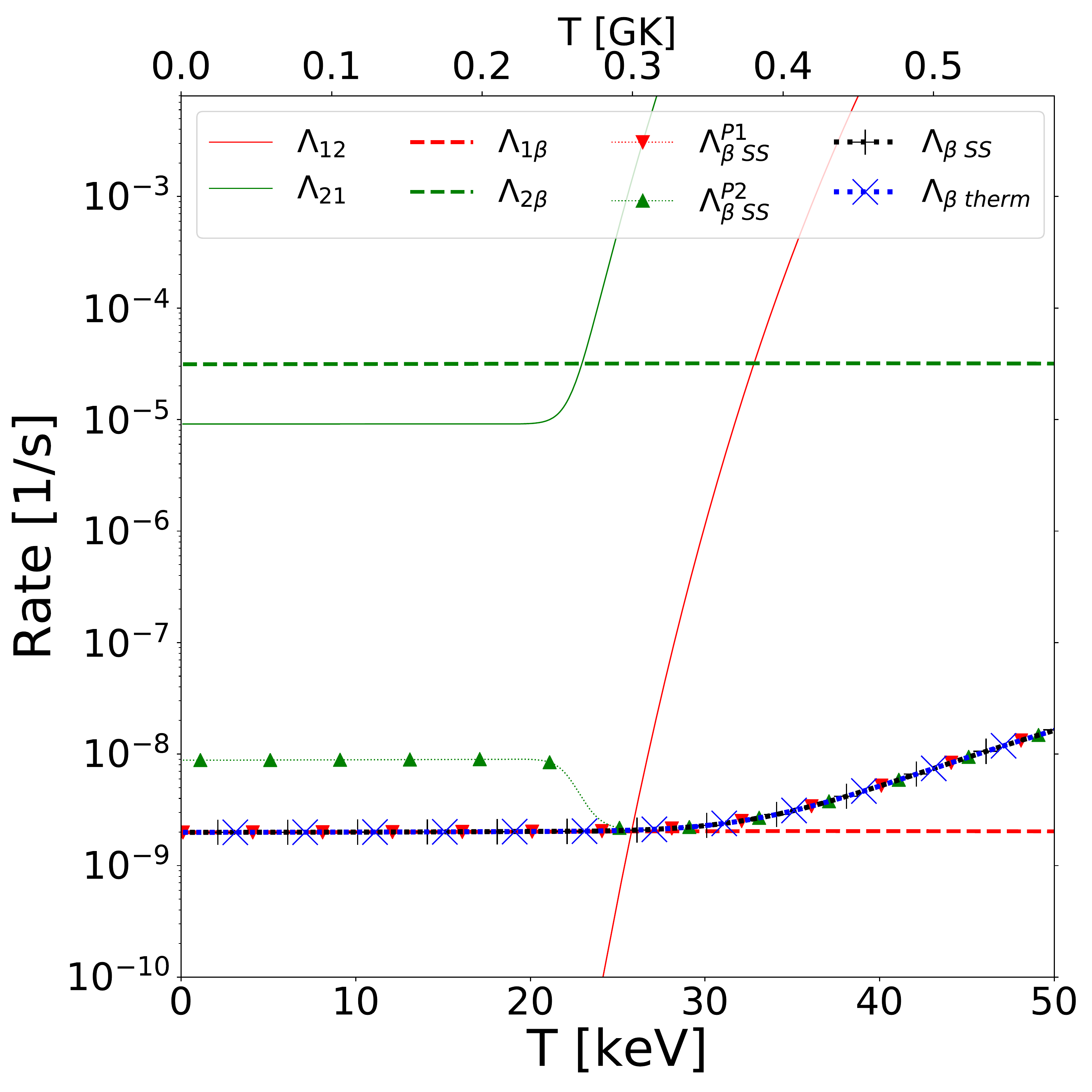}
    \caption{\Kr{} transition and $\beta$-decay rates.  The lines are as in figure \ref{fig:al_lambdas}.}
    \label{fig:kr_rates}
\end{figure}

We will not address the neutron capture cross section, but we did apply our method to study the isomer's effect on $\beta$ decay.  We used experimental data on the lowest 30 levels of {\Kr}.  We supplemented unmeasured transition rates with the Weisskopf approximation; {\Kr} ($N=49$) is near a closed neutron shell ($N=50$)---which diminishes the effectiveness of shell model calculations---so we made no further data supplements.  Figure \ref{fig:kr_rates} shows our results.

For the most part, {\Kr} doesn't behave in any remarkable way; the thermal and steady-state decay rates track each other, including when {\Kr} is produced in the ground state, so the isomer is not an astromer.  This is because in contrast with {\Al}, $\Lambda_{21}$ exceeds $\Lambda_{2\beta}$ (the green lines cross) at a lower temperature than $\Lambda_{12}$ exceeds $\Lambda_{1\beta}$ (the red lines cross).  Therefore, by the time the ground state ensemble can appreciably feed it, the isomer ensemble preferentially transitions back to ground rather than $\beta$ decays, and the thermal $\beta$-decay rate is appropriate for all situations except when {\Kr} is produced in the isomer ensemble.

The latter case, however, can give rise to interesting behavior.  Below $T\approx 21$ keV, material produced in the isomer ensemble has an $\sim 80\%$ chance to $\beta$ decay rather than transition to ground; this drives up the decay rate of the species and favors $\beta$ decay over neutron capture, rendering the {\Kr} isomer an astromer.  Above $T\approx 25$ keV, the nuclear levels thermalize, and the thermal $\beta$-decay rate is always correct.  The weak component of the $s$-process occurs in massive stars with temperatures between 30 and 90 keV, well above the {\Kr} thermalization temperature.  But interestingly, the main $s$-process occurs in pulsing asymptotic giant branch stars of mass $1-3$ M$_\odot$; short pulses reach temperature $T\sim 30$ keV, while longer interpulse periods have a temperature of $T\sim 8$ keV.  The entire cycle has a period of $\sim 50$ thousand years \citep{busso1999nucleosynthesis}.  This means that the {\Kr} isomer alternates being an astromer in the main $s$-process.

The uncertainties in the {\Kr} effective transition rates are somewhat different from the {\Al} and {\Cl} cases in two ways.  First, the experimental uncertainties are asymmetric (see table \ref{tab:85kr_sensitivity}).  Second, the unmeasured rates are estimated with the Weisskopf approximation rather than the shell model; the Weisskopf approximation tends to only be precise to one or two orders of magnitude.

\begin{figure}
    \centering
    \includegraphics[width=\columnwidth]{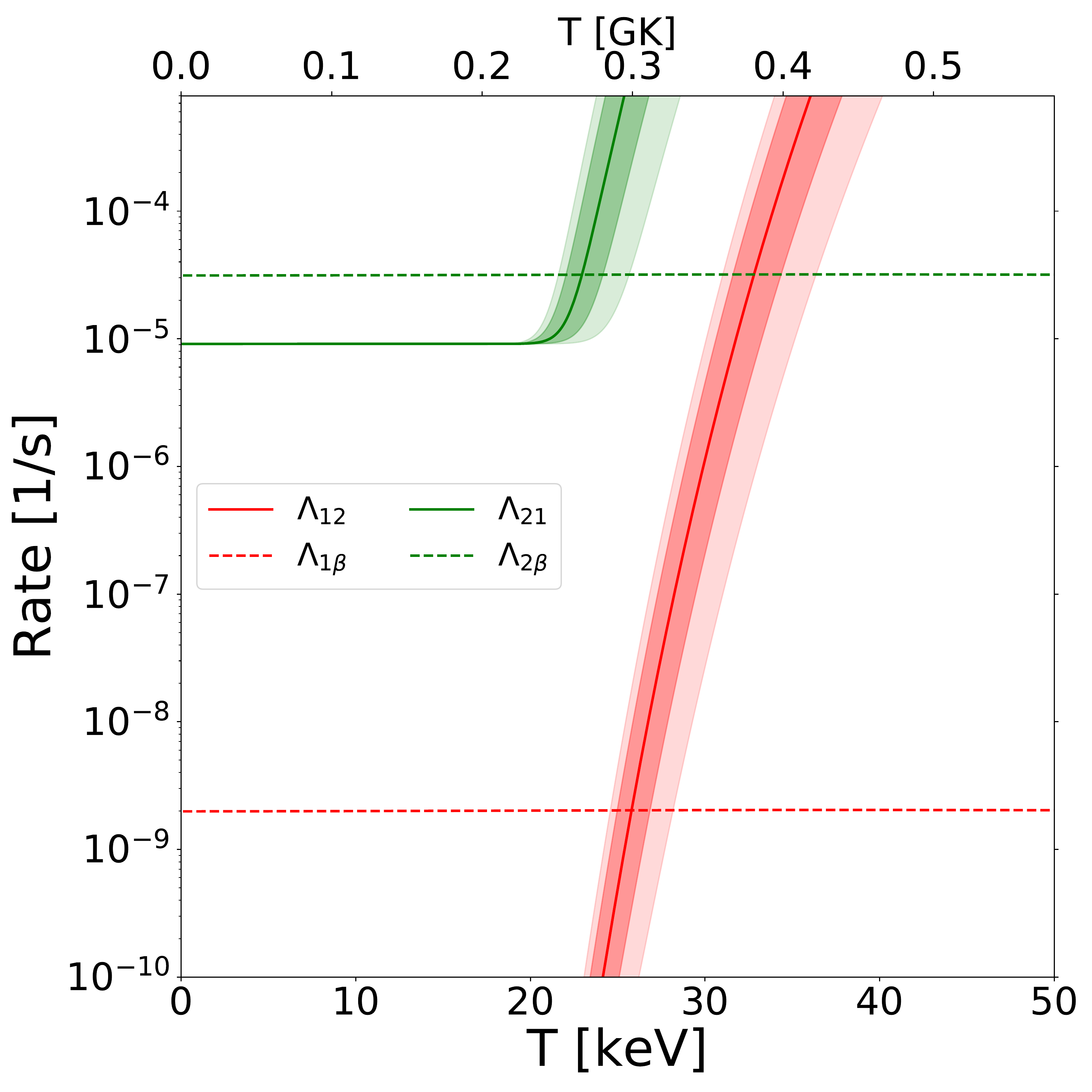}
    \caption{Range of {\Kr} effective transition rates.  Dark bands: measured rates increased/decreased by one standard deviation, Weisskopf rates multiplied/divided by a factor of 10.  Light bands: measured rates increased/decreased by two standard deviations, Weisskopf rates multiplied/divided by a factor of 100.}
    \label{fig:85kr_uncertainty}
\end{figure}

Figure \ref{fig:85kr_uncertainty} shows the uncertainties in the \nuc{85}{Kr} $\Lambda_{AB}$.  The upper light band is much narrower than the lower light band due to a bottle-necking effect similar to \nuc{34}{Cl}.  As the $\lambda_{st}$ are turned up together, the experimental $\lambda_{41}$ becomes a bottleneck; because its uncertainty is much less than the Weisskopf rates, it is turned up less than the Weisskopf rates and eventually controls the $\Lambda_{AB}$.  However, as the rates are turned down, the opposite happens: $\lambda_{41}$ is reduced much more slowly, the Weisskopf rates become the bottleneck, and the lower band is wider than the upper band.  If we decrease the Weisskopf rates much more, then as with \nuc{34}{Cl}, the $1 \rightarrow 2$ direct transition will be favored up to higher temperatures, and the $\Lambda_{AB}$ will be insensitive to further decreases.

We show the sensitivity of the $\Lambda_{AB}$ to selected $\lambda_{st}$ in table \ref{tab:85kr_sensitivity}; we used the same selection criteria as with {\Al}.  Similarly, we only vary by one ``unit'' of uncertainty, which will avoid the bottleneck effect.  We examine the fairly narrow range of temperatures from 21-25 keV because that is the range of thermalization temperatures found from figure \ref{fig:85kr_uncertainty}.  At all temperatures, the unmeasured $\lambda_{43}$ is the largest contributor to the effective transition rate uncertainty.  At $T=23$ keV, the unmeasured $\lambda_{54}$ plays a significant role, and at $T=25$ keV, the unmeasured $\lambda_{65}$ and the measured (but uncertain) $\lambda_{41}$ are also important sources of uncertainty in the $\Lambda_{AB}$.

\begin{table}
    \centering
    \caption{Most-uncertain individual transitions in the dominant pathways through {\Kr}.  The type column indicates whether this spontaneous transition rate comes from experiment (Exp) or the Weisskopf approximation (W).  The column labeled Fraction shows what fraction of the effective transitions flow through paths containing the individual transition (or its reverse).  Variation is an estimate of the uncertainty; we use the published uncertainty for experimental rates and a factor of 10 for Weisskopf rates.  The last column shows the fractional change in effective transition rates when the individual rate is adjusted up and down by one ``unit'' of uncertainty.}
    \begin{tabular}{c|c|c|c|c|c}
        T & Trans & Type & Fraction & Variation & Impact \\
        \hline\hline
        21 & $3\rightarrow 2$ & W & 0.0605 & $\times 10$ & 0.9995--1.0001 \\
         & $4\rightarrow 1$ & Exp & 0.0698 & -40,+80\% & 0.9901--1.0085 \\
         & $4\rightarrow 3$ & W & 0.0605 & $\times 10$ & 0.9492--1.1517 \\
         & $5\rightarrow 2$ & W & 0.0047 & $\times 10$ & 1.0 \\
         & $5\rightarrow 4$ & W & 0.0047 & $\times 10$ & 0.9965--1.0301 \\
         & $6\rightarrow 2$ & Exp & 0.0047 & -17,+25\% & 0.9995--1.0007 \\
         & $6\rightarrow 4$ & W & 0.0084 & $\times 10$ & 0.9962--1.0015 \\
        \hline
        23 & $3\rightarrow 2$ & W & 0.5725 & $\times 10$ & 0.9941--1.0006 \\
         & $4\rightarrow 1$ & Exp & 0.6797 & -40,+80\% & 0.8964--1.0907 \\
         & $4\rightarrow 3$ & W & 0.5725 & $\times 10$ & 0.5248--2.3960 \\
         & $5\rightarrow 2$ & W & 0.0487 & $\times 10$ & 1.0 \\
         & $5\rightarrow 4$ & W & 0.0487 & $\times 10$ & 0.9633--1.3098 \\
         & $6\rightarrow 2$ & Exp & 0.0585 & -17,+25\% & 0.9938--1.0087 \\
         & $6\rightarrow 4$ & W & 0.1051 & $\times 10$ & 0.9506--1.0200 \\
        \hline
        25 & $3\rightarrow 2$ & W & 0.7312 & $\times 10$ & 0.9914--1.0009 \\
         & $4\rightarrow 1$ & Exp & 0.9219 & -40,+80\% & 0.8526--1.1312 \\
         & $4\rightarrow 3$ & W & 0.7312 & $\times 10$ & 0.4008--2.7348 \\
         & $5\rightarrow 2$ & W & 0.0971 & $\times 10$ & 1.0 \\
         & $5\rightarrow 4$ & W & 0.0671 & $\times 10$ & 0.9490--1.4223 \\
         & $6\rightarrow 2$ & Exp & 0.0936 & -17,+25\% & 0.9900--1.0139 \\
         & $6\rightarrow 4$ & W & 0.1979 & $\times 10$ & 0.9174--1.0335 \\
         & $6\rightarrow 5$ & W & 0.0300 & $\times 10$ & 0.9822--1.1087
    \end{tabular}
    \label{tab:85kr_sensitivity}
\end{table}

\subsection{Other $s$-process Astromers}
\label{sec:s-process}

To reiterate from section \ref{sec:kr}, the main $s$-process occurs in thermally-pulsing asymptotic giant branch (AGB) stars.  The total pulsation period is $\sim 50$ thousand years, with a pulse temperature of $T \sim 30$ keV and an interpulse temperature of $T \sim 8$ keV \citep{busso1999nucleosynthesis}.

All rates in this section were computed using all measured levels in each nucleus (up to 30 levels) with unmeasured $\gamma$-decay rates computed from the Weisskopf approximation.  We do not provide sensitivity studies for these nuclei because nearly all of the intermediate level lifetimes (and hence $\gamma$ rates) along dominant transition paths are unknown, and the uncertainties in the $\Lambda_{AB}$ are driven by the Weisskopf approximation.

\begin{figure}
    \centering
    \includegraphics[width=\columnwidth]{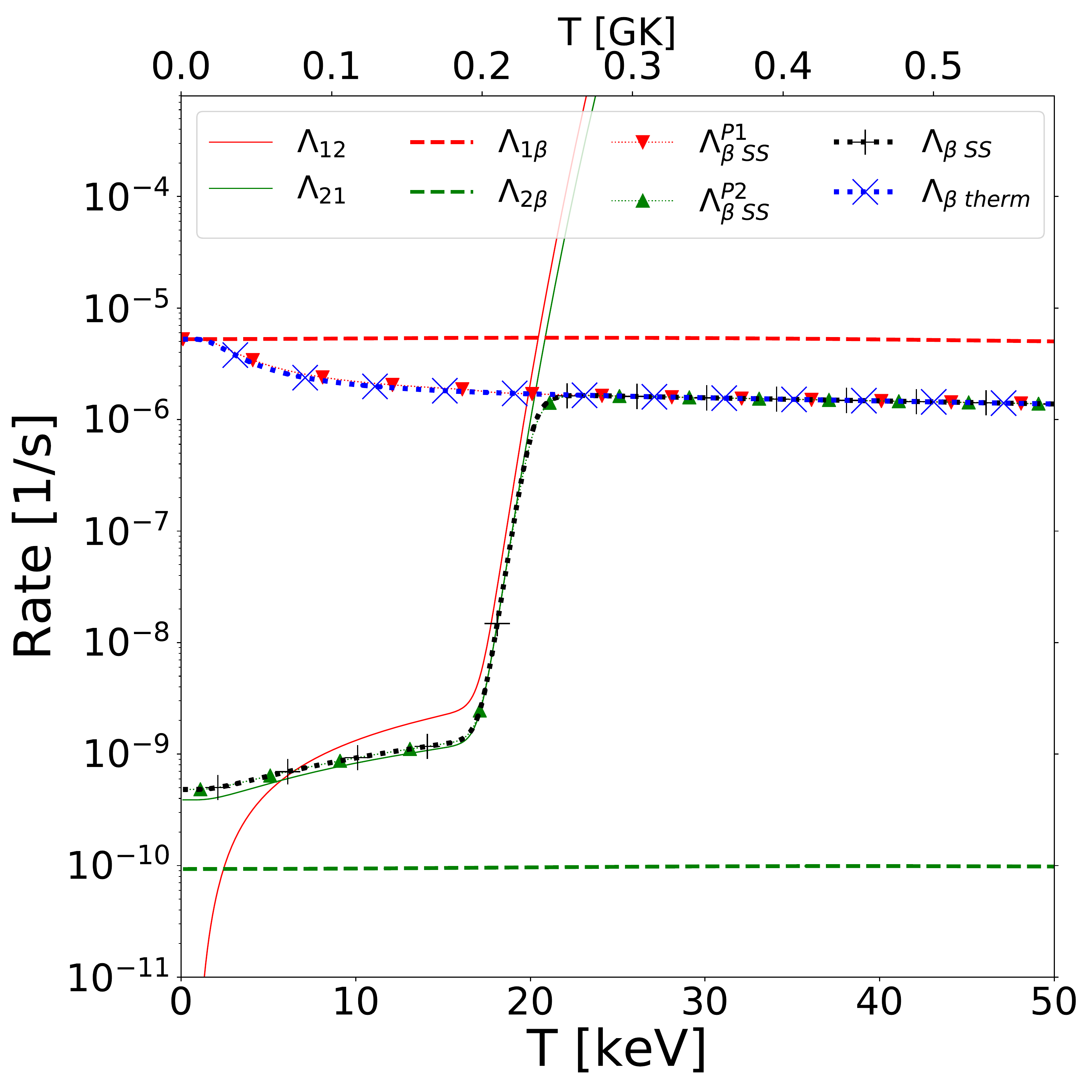}
    \caption{\nuc{121}{Sn} transition and $\beta$-decay rates.  The lines are as in figure \ref{fig:al_lambdas}.}
    \label{fig:121sn_rates}
\end{figure}

The 6.31 keV isomer of \nuc{121}{Sn} ($T_{1/2} = 43.9$ y) is more stable than the ground state ($T_{1/2} = 27.03$ h).  Unlike other isomers considered in this work, this one decays primarily to the ground state (77.6\%) rather than $\beta$ decay, so $\Lambda_{21}$ always exceeds $\Lambda_{2\beta}$.  However, figure \ref{fig:121sn_rates} shows that $\Lambda_{12}$ doesn't catch up to $\Lambda_{1\beta}$ until temperature $T \sim 20$ keV, making this an astromer in the interpulse periods of the main $s$-process.  Moreover, the long half-life of the astromer provides ample time for it to capture a neutron rather than de-excite or $\beta$ decay.  This would reduce the $s$-process production of \nuc{121}{Sb}, \nuc{122}{Te}, and \nuc{123}{Te}.

\begin{figure}
    \centering
    \includegraphics[width=\columnwidth]{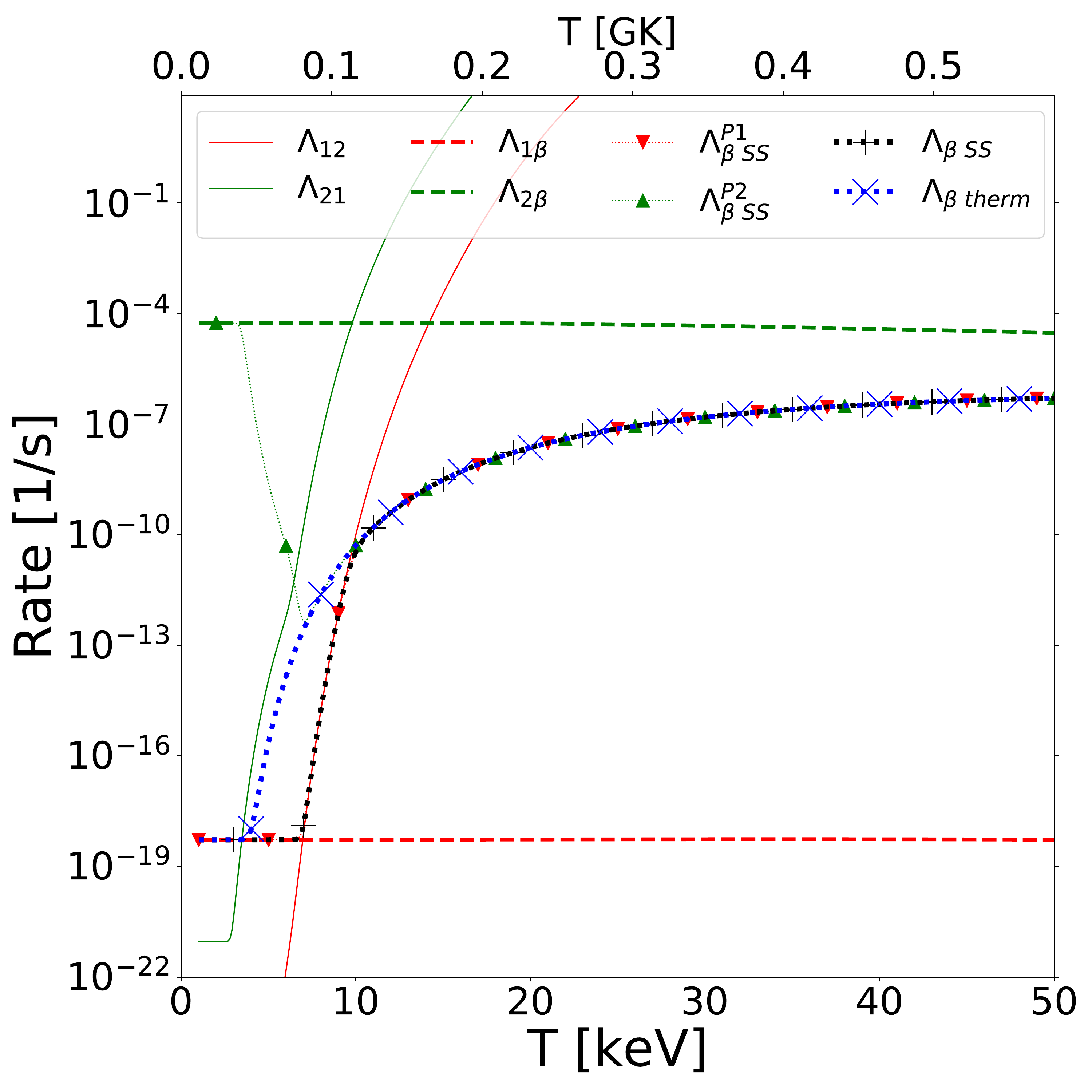}
    \caption{\nuc{176}{Lu} transition and $\beta$-decay rates.  The lines are as in figure \ref{fig:al_lambdas}.}
    \label{fig:176lu_rates}
\end{figure}

\cite{takahashi1987beta} took advantage of the 122.845 keV isomer to compute a thermal $\beta$-decay rate for \nuc{176}{Lu}.  We also computed the transition and $\beta$-decay rates in this isotope; the results are in figure \ref{fig:176lu_rates}.  \cite{takahashi1987beta} performed a very careful calculation, but two of their temperature points (0.05 and 0.1 GK) fall in the range where the two states should be treated as separate species.  Furthermore, \nuc{176}{Lu} is made in the main $s$-process, and the $\sim 10$ keV thermalization temperature implies its isomer could be an astromer during the main $s$-process interpulse periods.

\subsection{$r$-process Astromers}
\label{sec:$r$-process}

All rates in this section were computed using all measured levels in each nucleus (up to 30 levels) with unmeasured $\gamma$-decay rates computed from the Weisskopf approximation.  We do not provide sensitivity studies for these nuclei because nearly all of the intermediate level lifetimes (and hence $\gamma$ rates) are unknown, and the uncertainties in the $\Lambda_{AB}$ are driven entirely by the Weisskopf approximation.

\begin{figure}
    \centering
    \includegraphics[width=\columnwidth]{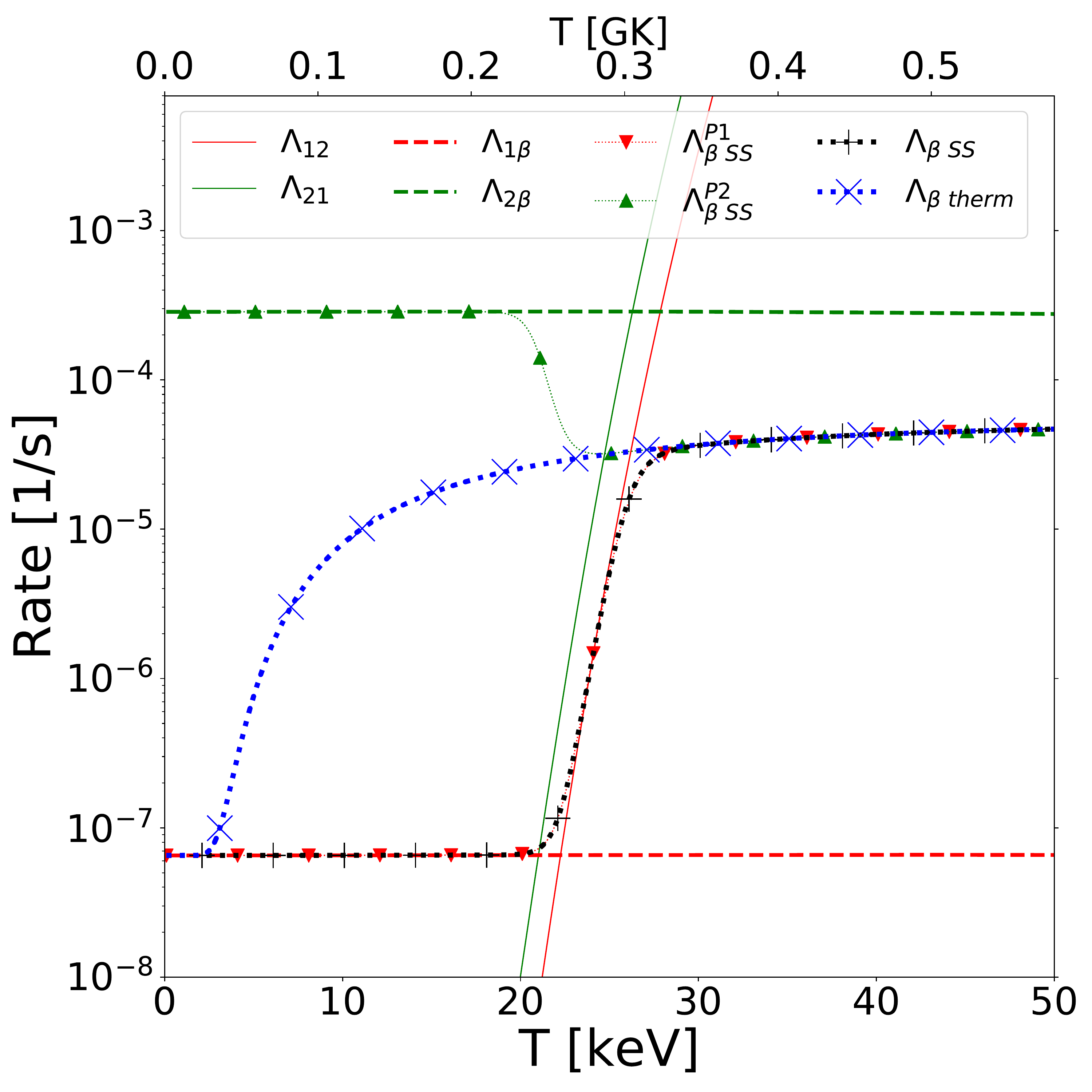}
    \caption{\nuc{123}{Sn} transition and $\beta$-decay rates.  The lines are as in figure \ref{fig:al_lambdas}.}
    \label{fig:123sn_rates}
\end{figure}

\begin{figure}
    \centering
    \includegraphics[width=\columnwidth]{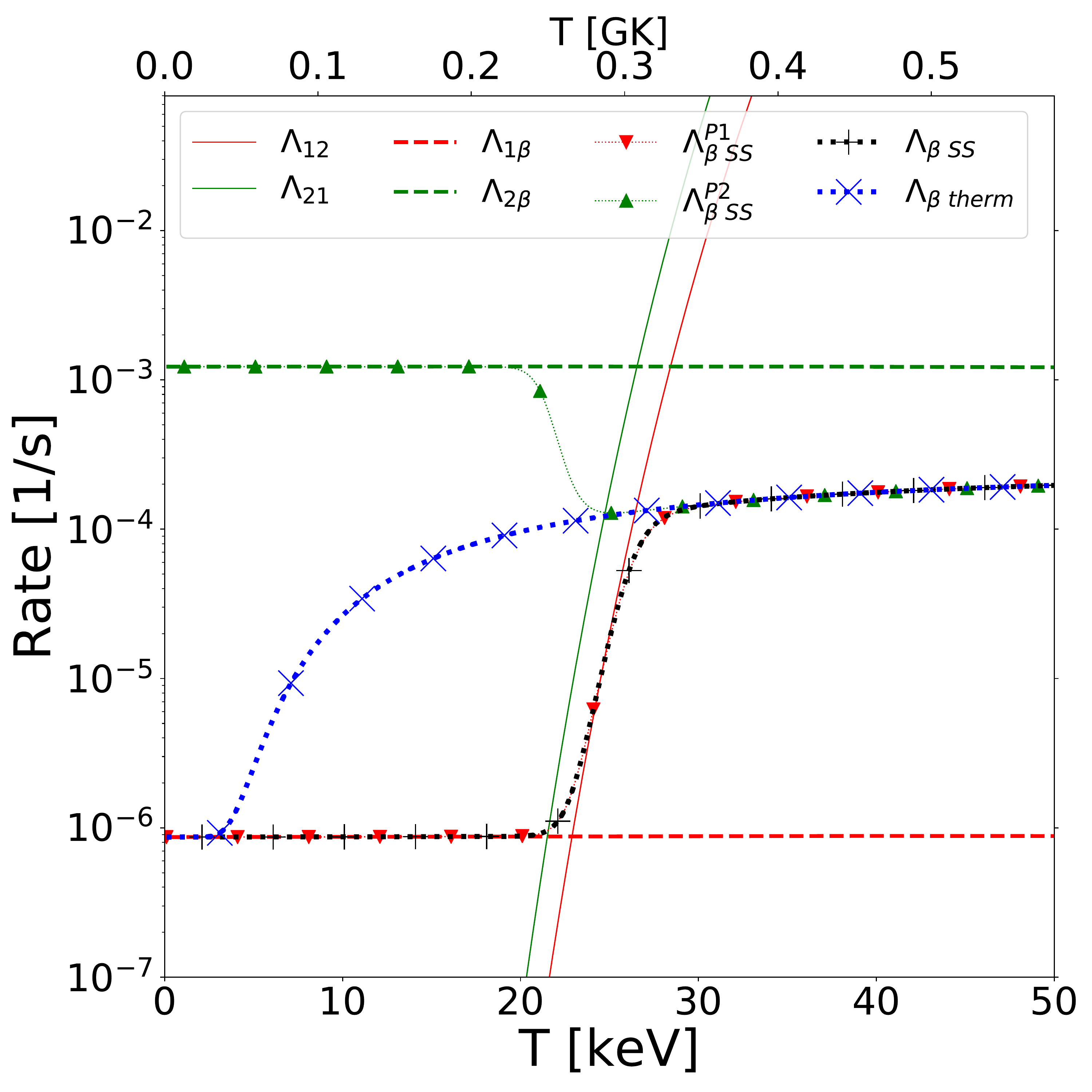}
    \caption{\nuc{125}{Sn} transition and $\beta$-decay rates.  The lines are as in figure \ref{fig:al_lambdas}.}
    \label{fig:125sn_rates}
\end{figure}

\begin{figure}
    \centering
    \includegraphics[width=\columnwidth]{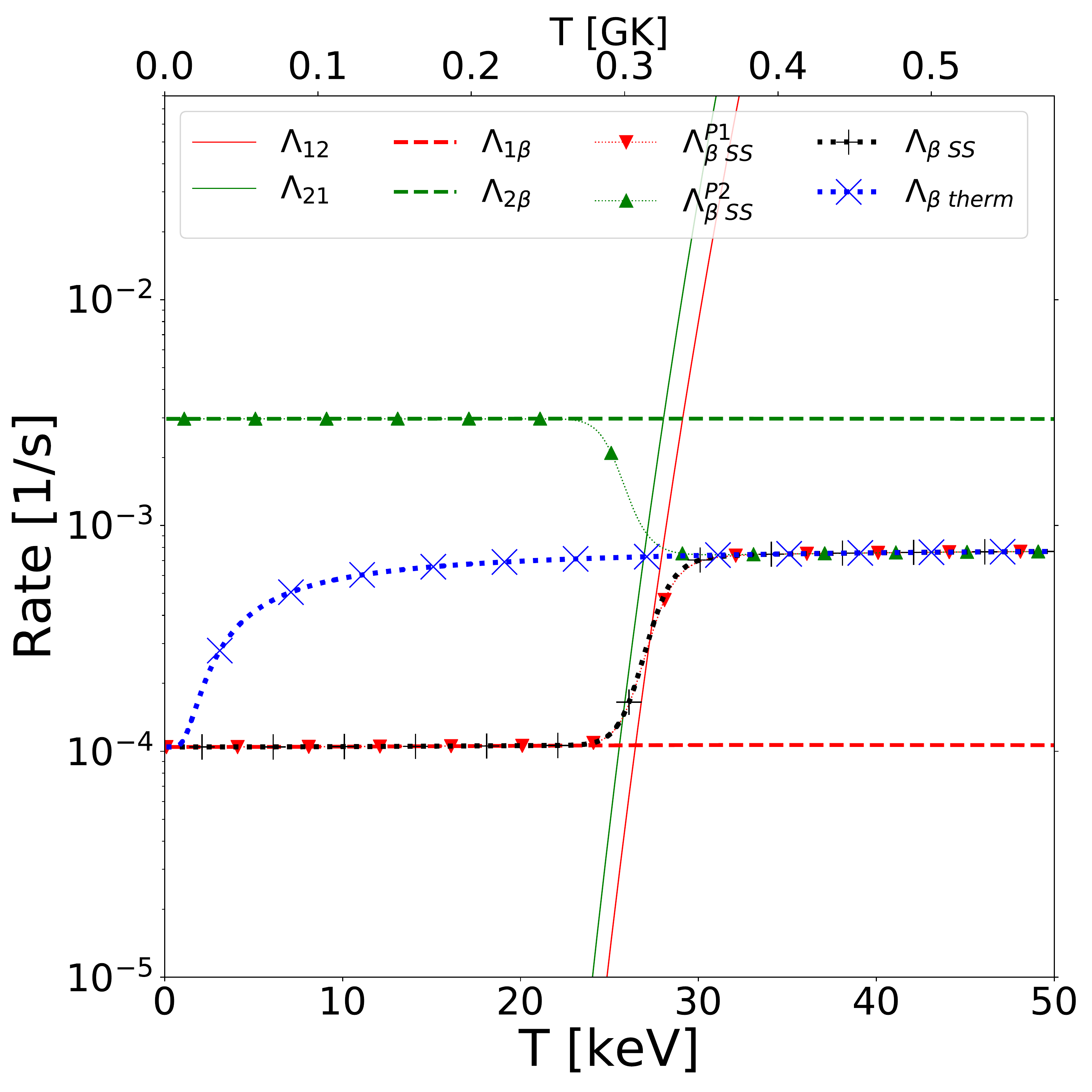}
    \caption{\nuc{127}{Sn} transition and $\beta$-decay rates.  The lines are as in figure \ref{fig:al_lambdas}.}
    \label{fig:127sn_rates}
\end{figure}

Isomers may play an important role in the rapid neutron capture process ($r$-process).  \cite{fujimoto2020impact} identified nine nuclei with isomers that could affect the light curve of a post-neutron-star-merger kilonova, four of which were particularly impactful: \nuc{123,125,127}{Sn} with isomers at 24.6, 27.50, and 5.07 keV respectively, and \nuc{128}{Sb} with an isomer of unknown energy.  Figures \ref{fig:123sn_rates}, \ref{fig:125sn_rates}, and \ref{fig:127sn_rates} show the transition and $\beta$-decay rates for the Sn isotopes.  Indeed, each of these thermalizes at $T \approx 25$ keV, well above the expected ambient temperature when the isotopes are populated in such environments.  The disparity between the ground state and isomer $\beta$-decay rates could have a marked impact on the radioactive heating as nuclei decay back to stability, so further investigation is in order.

\begin{figure}
    \centering
    \includegraphics[width=\columnwidth]{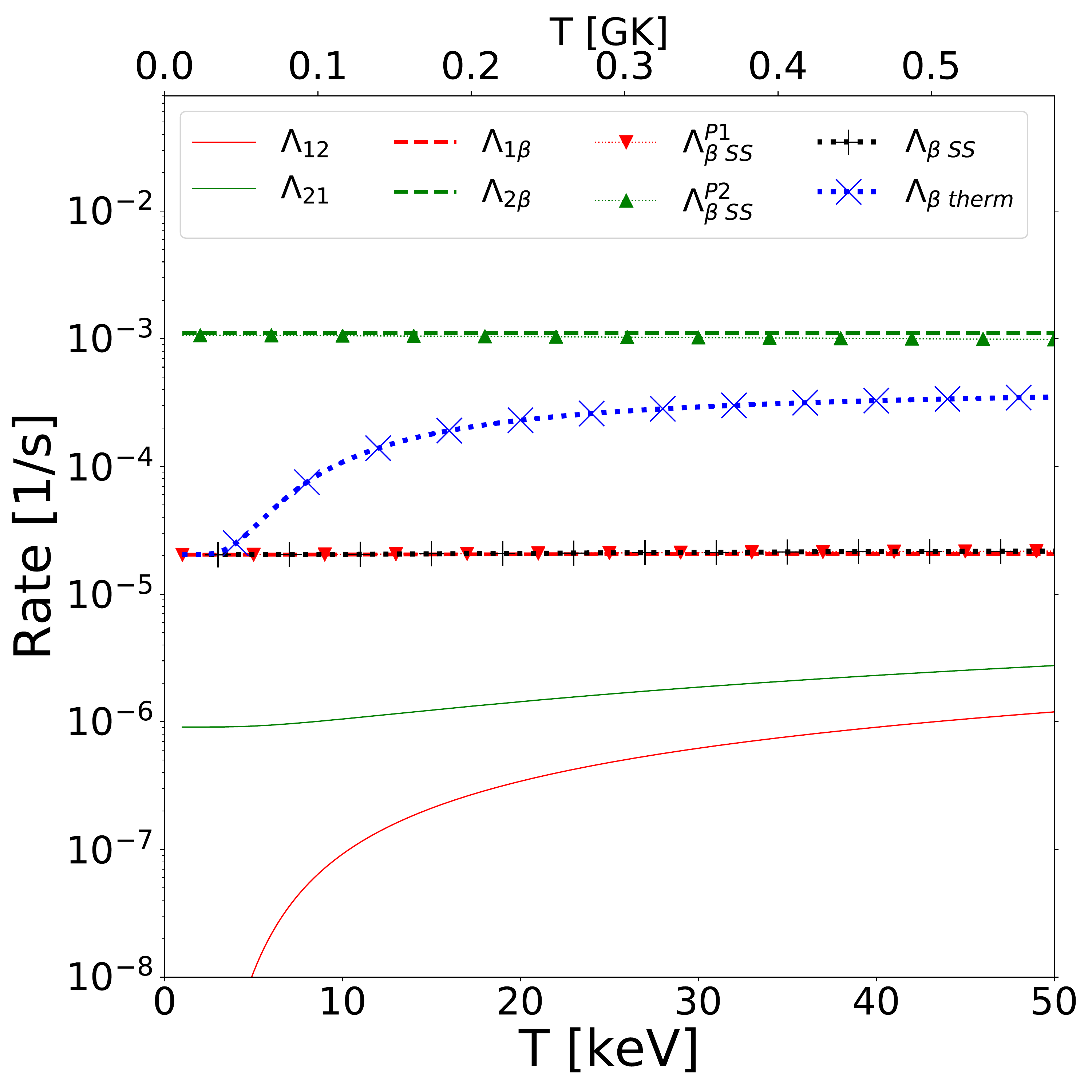}
    \caption{\nuc{128}{Sb} transition and $\beta$-decay rates.  The lines are as in figure \ref{fig:al_lambdas}.}
    \label{fig:128sb_rates}
\end{figure}

The fourth isotope identified by \cite{fujimoto2020impact}, \nuc{128}{Sb}, suffers from incomplete data.  The rates are shown in figure \ref{fig:128sb_rates}, but it does not thermalize within the computed temperature range.  This is almost certainly due to missing data in the excited states.  The ground state has spin and parity $J^\pi = 8^-$, the isomer has $J^\pi = 5^+$, and all other measured states have $J <= 4$.  Furthermore, the isomer has unknown energy (we take the likely upper bound of 20 keV), and all other excited states have energies measured with respect to that unknown energy \citep{elekes2015nuclear}.  Because this is such a potentially important nucleus, closer investigation is warranted.  If the isomer is indeed influential, this would motivate experiments to measure the excited state properties.

\begin{figure}
    \centering
    \includegraphics[width=\columnwidth]{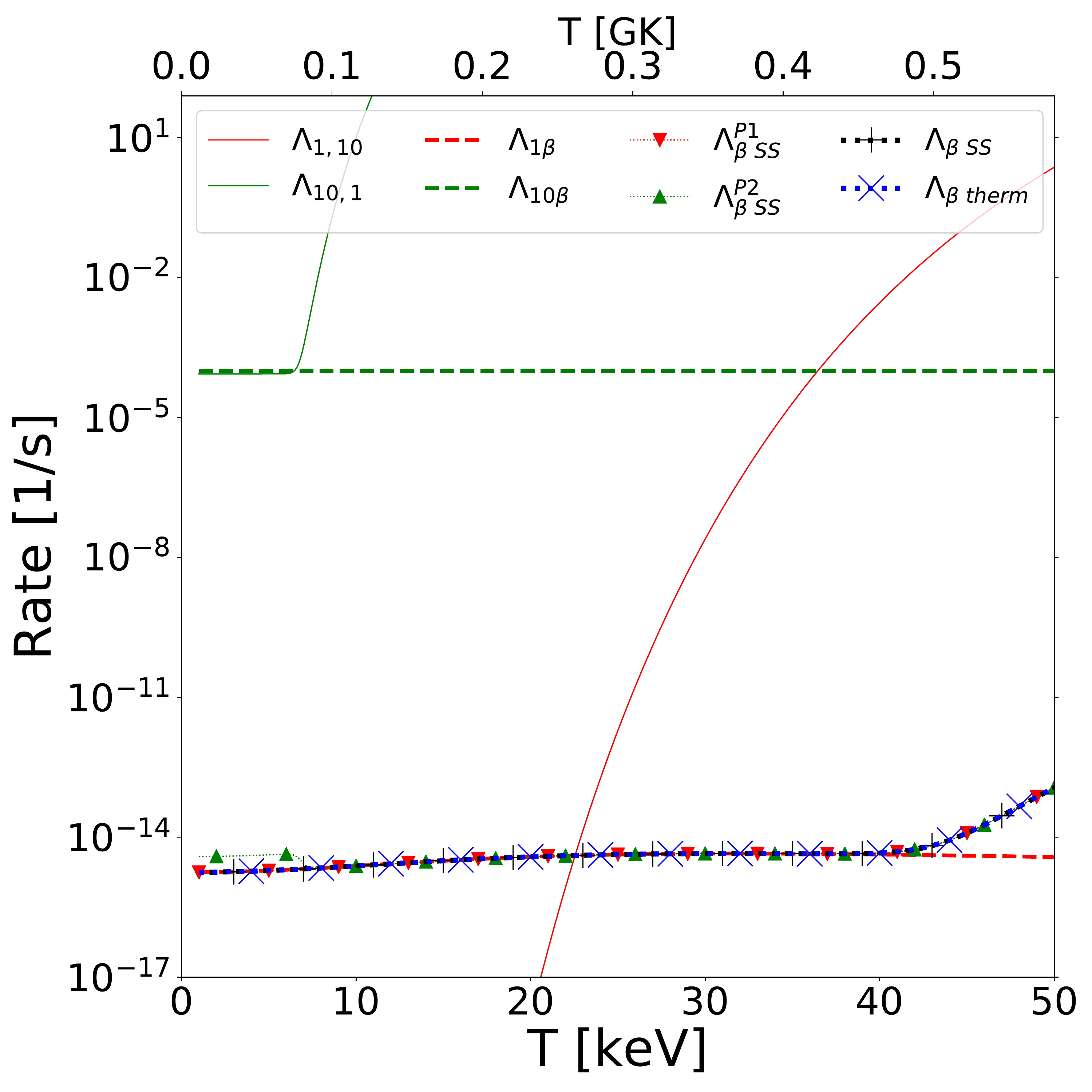}
    \caption{\nuc{182}{Hf} transition and $\beta$-decay rates.  The lines are as in figure \ref{fig:al_lambdas}.}
    \label{fig:182hf_rates}
\end{figure}

\nuc{182}{Hf} is a cosmochronometer that may be produced in the $s$-process \citep{lugaro2014stellar} or the $r$-process \citep{wu2019finding}.  It has an isomer at 1172.87 keV; figure \ref{fig:182hf_rates} shows the rates for this isotope.  Although \nuc{182}{Hf} thermalizes at a fairly low $T \approx 6$ keV and is therefore not an $s$-process astromer, this is nevertheless higher than the expected temperature at which it would be produced in some $r$-process events; the two-minute $\beta$-decay half life of \nuc{182}{Lu} likely implies that a rapid neutron capture event would have adequate time to cool.  However, \nuc{182}{Lu} $\beta$ decay exclusively feeds the \nuc{182}{Hf} ground state ensemble \citep{kirchner1982new}.  With this information we conclude that the \nuc{182}{Hf} isomer does not affect the $\gamma$-ray signal predicted by \cite{wu2019finding}.

Finally, we raise the interesting case of \nuc{170}{Ho}.  This nuclide has only two measured levels: the ground state with $J^\pi = 6^+$ and a $\beta$-decay half life of 2.76 minutes, and a 120 keV isomer with $J^\pi = 1^+$ and a $\beta$-decay half life of 43 seconds.  We anticipate that this difference in decay rates could affect energy generation in an $r$-process event with observable effects in the light curve of a kilonova.  \nuc{170}{Ho} is produced by $\beta$ decay of \nuc{170}{Dy}, but the parent $\beta$-decay intensities are unmeasured.  However, the parent ground state has $J^\pi = 0^+$, so it likely decays predominantly to the \nuc{170}{Ho} isomer.  This fact---coupled with the lack of data on \nuc{170}{Ho} excited states (which greatly hinders estimating the $\Lambda_{AB}$), the computed abundance of $A=170$ nuclei \citep{Sprouse2020, Vassh2020}, and the potential kilonova implications---motivates experimental examination of this pair of nuclei.

\subsection{Other Cases}
\label{sec:non-astromers}

We show here two examples of isomers which look promising as astromers but for which we are unaware of any environments where they may behave as such.  We computed unmeasured $\gamma$-decay rates using the Weisskopf approximation.

\begin{figure}
    \centering
    \includegraphics[width=\columnwidth]{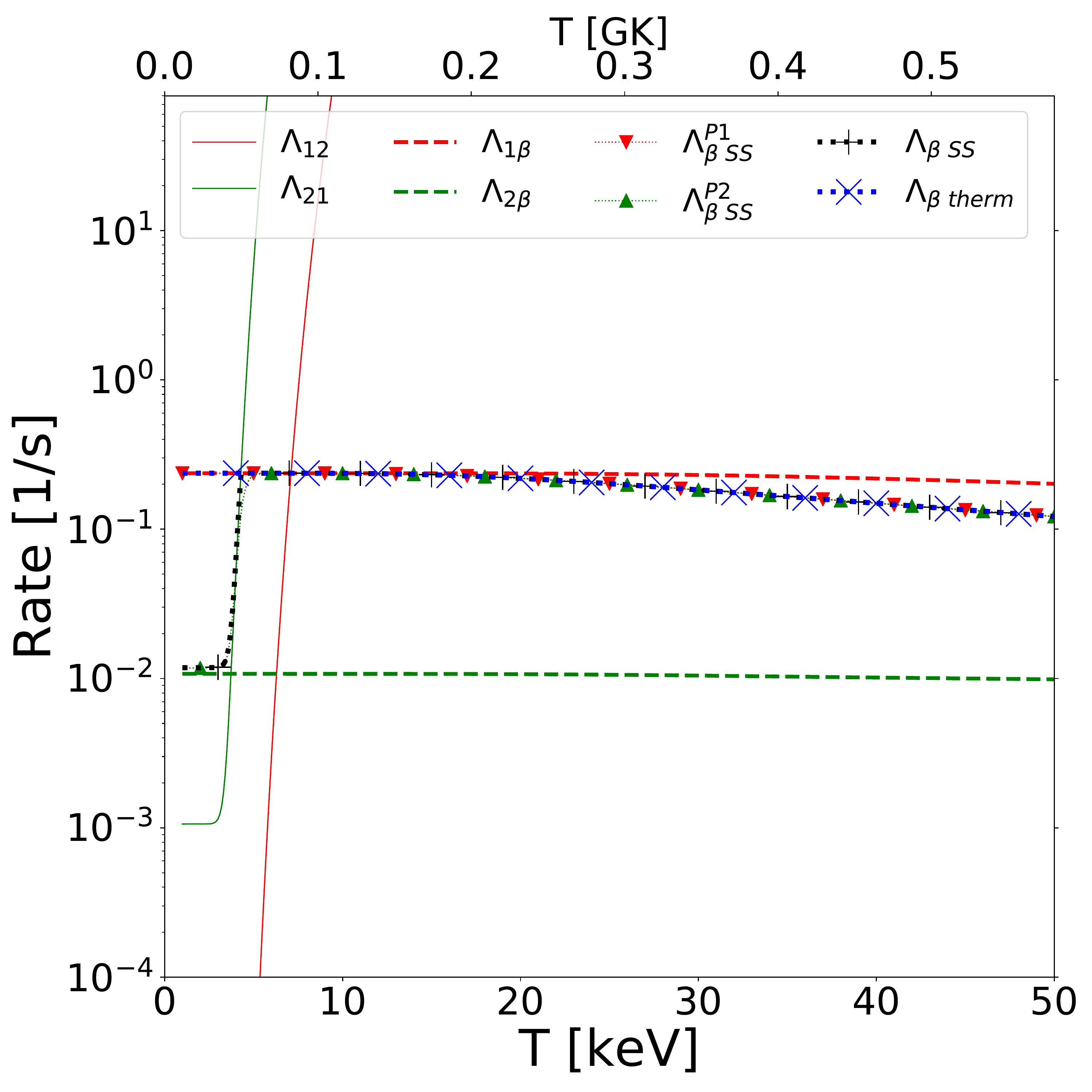}
    \caption{\nuc{58}{Mn} transition and $\beta$-decay rates.  The lines are as in figure \ref{fig:al_lambdas}.}
    \label{fig:58mn_rates}
\end{figure}

\nuc{58}{Mn} is an important antineutrino source in pre-supernova stellar cores \citep{patton2017neutrinos}, and it has an isomer at 71.77 keV.  We computed effective transition and $\beta$-decay rates using the lowest 30 measured levels.  As shown in figure \ref{fig:58mn_rates}, \nuc{58}{Mn} thermalizes at $T<5$ keV, far below the pre-supernova core temperatures of $300-900$ keV, so it is not an astromer in this environment.  Nevertheless, there is a point to be made about this isomer.  Because \nuc{58}{Mn} is such a prodigious antineutrino source (with consequences for pre-supernova detection),  we naturally desire precise weak interaction rates.  In most cases, we must rely entirely on theory for the weak interaction strengths of excited states, but the existence of a low-lying isomer gives precise experimental data that reduces uncertainties in the high-temperature weak interaction rates of \nuc{58}{Mn}.

\begin{figure}
    \centering
    \includegraphics[width=\columnwidth]{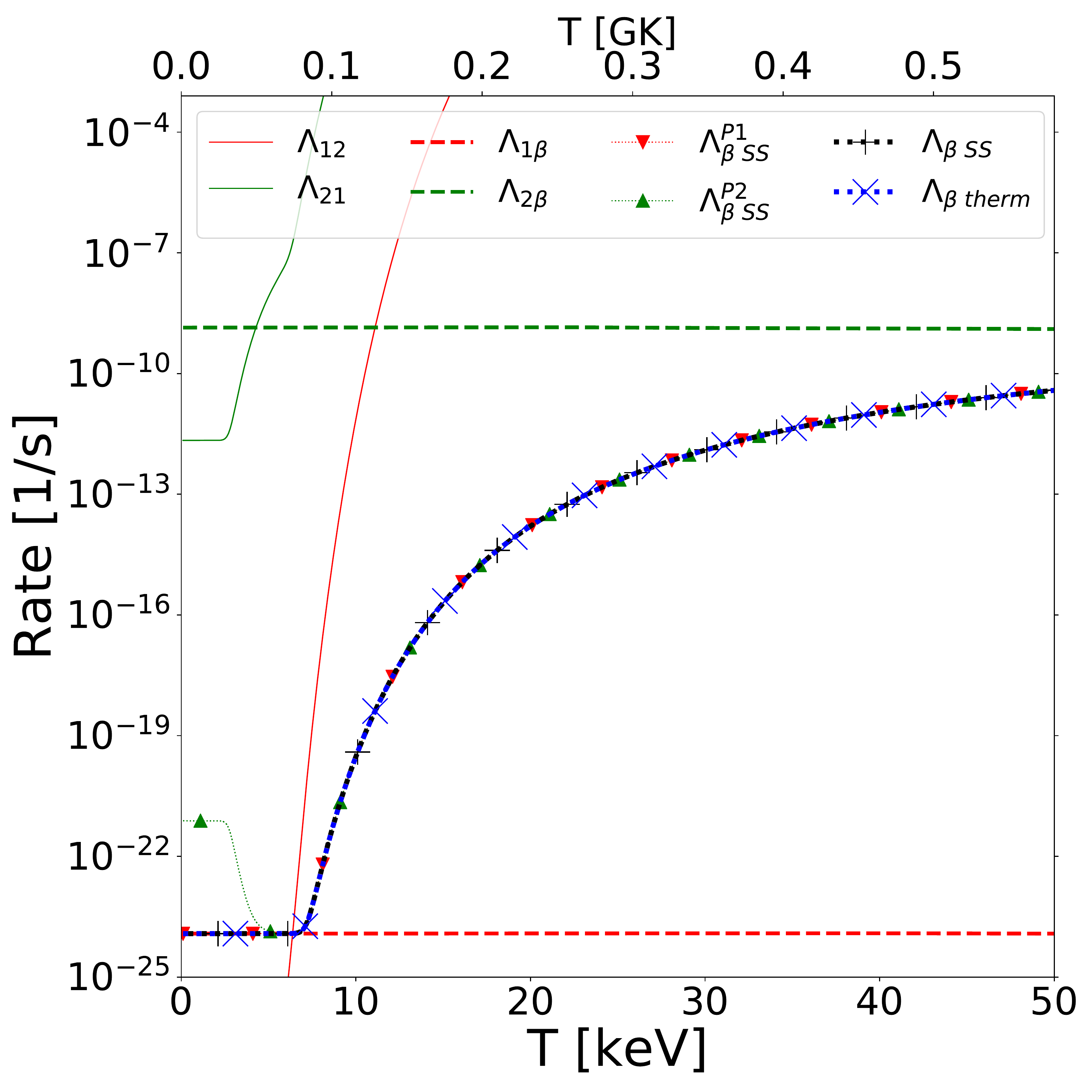}
    \caption{\Cd{} transition and $\beta$-decay rates.  The lines are as in figure \ref{fig:al_lambdas}.}
    \label{fig:cd_rates}
\end{figure}

{\Cd} has a 14.1 year isomer at 263.54 keV that nearly always $\beta$ decays \citep{blachot2010nuclear}.  Because it lies on the $s$-process path and the isomer $\beta$-decay rate could make it a branch point, it is worth taking a closer look at.  We computed effective transition rates and $\beta$-decay rates in {\Cd} using the 30 lowest measured energy levels; our results are shown in figure \ref{fig:cd_rates}.

The {\Cd} rates are qualitatively similar to {\Kr} in that $\Lambda_{21}$ overtakes $\Lambda_{2\beta}$ at a lower temperature than $\Lambda_{12}$ exceeds $\Lambda_{1\beta}$ with similar consequences for the $\beta$-decay rates.  The major difference is that the thermalization temperature for {\Cd} is only $\sim 5$ keV.  This is below any $s$-process temperatures, so this isomer is likely never an $s$-process astromer.

Of course, as with \nuc{58}{Mn}, the existence of the isomer means that we have a precise measurement of the $\beta$-decay rate for the first excited state in this nucleus, providing a more precise value of the thermal $\beta$-decay rate for this species.  Also, if there exists an environment where the isomer is produced at a cooler temperature, {\Cd} that is produced in the ground state will remain nearly stable, but that produced in the isomer will decay to $^{113}$In.  However, we are unaware of any specific astrophysical environments where this would play a role, since \nuc{113}{Ag}---the {\Cd} $\beta$ decay parent produced in the $r$-process---essentially always decays to the ground state ensemble.

\section{Discussion and Conclusions}
\label{sec:conclusions}

We have developed an effective means of computing thermally driven transition rates between the ground state and long-lived isomers in atomic nuclei.  We focus on the case of a single isomer, but generalization to multiple isomers is straightforward by simply including more endpoint states that obey all of our well-defined rules.  Our technique does not rely on any physical assumptions apart from the assumption of a thermal photon bath, and, as discussed below, even this can be relaxed.  While the technique itself is uncomplicated, we have extensively and rigorously proved its validity, including developing a detailed means of analyzing which transitions are important at a given temperature.

Although we developed our method independently from \cite{gupta-meyer:2001}, our computed rates will be numerically similar to calculations using the technique of that work because we ultimately arrive at similar conclusions.  Nevertheless, ours has some advantages.  First, whereas \cite{gupta-meyer:2001} assume that intermediate states equilibrate essentially instantaneously, we do not rely on physical assumptions.  Second, we do not use a Taylor expansion to solve equation \ref{eq:prob_vec}; a solution always exists, and a linear equation solver will quickly find it.  This eliminates any concern about including enough terms of the expansion for any nucleus at any temperature. 

We do, however, owe a great debt to \cite{gupta-meyer:2001} for our calculations of ensemble $\beta$-decay rates.  While these rely on the assumption of fast equilibration of intermediate states, it is in general a very good assumption, and their method of assigning ensemble weights to the intermediate states is a powerful way to accurately break the nucleus into two species for nucleosynthesis network calculations.

We have applied our methods to several interesting nuclei, and we used the formulation of \cite{gupta-meyer:2001} and \cite{mgs:2018} to compute $\beta$-decay rates for these nuclei, treating the ground state and isomers as separate species.  For most of our case studies, we find that for a given set of destruction rates (in this case, $\beta$ decay, though it applies to all destruction channels) there is a clear ``thermalization temperature''.  Above this temperature, the nuclear levels may be assumed to be in thermal equilibrium, and below it, the isotope should be treated as two independent species: the ground state and an astromer.  We summarize in Table \ref{tab:data_summary} the astrophysically relevant nuclear isomers studied in this work.

We have shown that the isomers in \nuc{121,123,125,127}{Sn}, \nuc{128}{Sb}, and \nuc{176}{Lu} have high thermalization temperatures relative to the environments where they are created.  Our results reinforce the claim by \cite{fujimoto2020impact} that some of these isomers play important roles in the $r$-process and show that there may be impact from astromers on the $s$-process.  We have also identified \nuc{170}{Ho} as another potentially impactful $r$-process astromer.  These findings motivate more detailed study of astromers in nucleosynthesis.

We find that there is currently insufficient data on the excited states of \nuc{128}{Sb} and \nuc{170}{Ho} to compute transition rates and thermalization temperatures.  This lack of data and their probable importance in the $r$-process highlights a need for experimental inquiry into their excited state properties.  Additionally, we have shown that uncertainties in individual intermediate state transitions can dramatically influence the effective transition rates and thermalization temperatures.  The success of precise measurements of {\Cl} transitions in producing precise effective transition rates demonstrates the value of measuring the unknown transitions in nuclei with isomers.

Our calculations have focused on $\beta$ decay, but the ensemble destruction rates can be just as readily computed for any other destruction channel.  We must simply recognize that a fast destruction rate will tend to raise the thermalization temperature.  For example, an isomer which decays solely via internal transition to the ground state will thermalize under most conditions.  But if the isomer has a very different neutron capture cross section from the ground state, it may fall out of equilibrium in the $r$-process, ultimately affecting nuclear flow.

Another concern is the ``feeding factors'' of the ground state and isomer during production; that is, for a given production channel and environment, what fraction of the material goes to which ensemble.  For example, consider the {\Al} rates shown in figure \ref{fig:al_lambdas}.  Assume for the sake of argument that we produce this nucleus at $T=10$ keV with a 50/50 split of ground state and isomer.  Before an appreciable population of the species has had a chance to build up, we would measure an instantaneous effective beta-decay rate for the species that is the average of the ground state and isomer decay rates.  However, as time goes on, a population of the ground state will build up while the isomer decays away.  When we again measure the instantaneous effective beta-decay rate, it will be skewed toward the ground state rate because the ground state will have a greater population.  We therefore see that the species effective beta-decay rate can be time-dependent irrespective of production mechanism.

Of course, the $P_{iE}$ for the states where the isotope is produced are the feeding factors into the ground and isomeric ensembles.  Consider {\Al} produced by \nuc{25}{Mg}(p,$\gamma$).  In this reaction, the initial states of {\Al} are a small handful of resonant levels above the proton separation energy of 6306.31 keV.  If we can reliably calculate the $P_{iE}$ for these states, we will know how much material lands in each of the isomer and ground state.  The same is true for the daughter levels in \nuc{84}{Kr}(n,$\gamma$){\Kr}, \nuc{170}{Dy}($\beta$)\nuc{170}{Ho}, and so on.

It's also important to note that while many of the finer details (path reversibility, etc.) rely on a thermal bath, the general method of computing effective isomer $\leftrightarrow$ ground state transitions does not.  As long as $\lambda_{st}$ and $\lambda_{ts}$ are measured or calculable for every relevant pair of states $s$ and $t$, we may use equation \ref{eq:prob_vec} to compute the effective transition rates.  This may have applications when nuclei with isomers are exposed to non-thermal sources of radiation.  In this vein, our method applies to random walks through any general weighted network of connected nodes where you wish to get from node A to node B without going through node C (or D, E, etc.)  Here the nodes are nuclear levels and the connections are internal nuclear transitions.

A nuclear isomer has astrophysical consequences and is hence an ``astromer'' below the thermalization temperature.  This temperature is sensitive to the various destruction rates that the nuclear species faces, with rapid rates increasing the temperature, thereby widening the range of conditions for which a metastable state is an astromer.  At sufficiently high temperatures, the transition rates dominate the destruction rates, the astromer property fades, and the isotope may be considered a single species.

\section{Supporting Data}

\begin{deluxetable*}{ccccccccccl}

\tablecaption{Summary of data for nuclei in this paper.  The column headers use $g$ and $m$ as shorthand for the ground state and isomer, respectively.  The state number (starting with ground $n_g=1$) of the isomer is $n_m=2$ for all nuclei except \nuc{182}{Hf}, which has $n_m=10$.  The isomer energy is $E_m$, and the $J^\pi$ are the spin and parity of the respective levels (parentheses denote uncertain $J^\pi$).  The half-lives ($T_{1/2}$) and $\beta$-decay branching for the isomer ($B_{m\beta}$) are as measured in the laboratory; $B_{m\beta}$ is the percent of isomer decays which are $\beta$ decays rather than internal transitions to another nuclear state.  The column \# States tells how many measured levels we included in our calculations, and our computed approximate thermalization temperature is indicated by $T_{\rm therm}$.  The Site column indicates the astrophysical site of interest.  Notes highlights important points about available data; specific states or ranges of states are deemed relevant from our pathfinding in that nucleus.
}

\tablehead{
\colhead{Isotope} & \colhead{$E_{m}$} & \colhead{$J^\pi_{g}$} & \colhead{$J^\pi_{m}$} & \colhead{$T_{1/2,~g}$} & \colhead{$T_{1/2,~m}$} & $B_{m\beta}$   & \# States  & $T_{\rm therm}$ & Site$^1$       & Notes \\ 
\colhead{}        & \colhead{(keV)}   & \colhead{}            & \colhead{}            & \colhead{(s)}          & \colhead{(s)}          & \colhead{(\%)} & \colhead{} & \colhead{(keV)} & \colhead{} & \colhead{}
}

\startdata
\nuc{26}{Al}  & 228.31  & $5^{+}$    & $0^{+}$    & $2.26\times 10^{13}$ & $6.35\times 10^{0}$  & 100   & 67 & 35      & $p$      & $\lambda_{32}$, $\lambda_{43}$ unmeasured \\
\nuc{34}{Cl}  & 146.36  & $0^{+}$    & $3^{+}$    & $1.53\times 10^{0}$  & $1.92\times 10^{3}$  & 55    & 30 & 20      & Sne      & $\lambda_{32}$ poorly constrained \\
\nuc{58}{Mn}  & 71.77   & $1^{+}$    & $4^{+}$    & $3.00\times 10^{0}$  & $6.54\times 10^{1}$  & 90    & 30 & 5       & PSne     & $\lambda_{ij}$ unmeasured for 3-7 \\
\nuc{85}{Kr}  & 304.87  & $9/2^{+}$  & $1/2^{-}$  & $3.39\times 10^{8}$  & $1.61\times 10^{4}$  & 78.8  & 30 & 25      & $s$      & $\lambda_{ij}$ poorly measured for 3-7 \\
\nuc{113}{Cd} & 263.54  & $1/2^{+}$  & $11/2^{-}$ & $2.54\times 10^{23}$ & $4.45\times 10^{8}$  & 99.86 & 30 & 5       & $r$      & $\lambda_{42}$ unmeasured \\
\nuc{121}{Sn} & 6.31    & $3/2^{+}$  & $11/2^{-}$ & $9.73\times 10^{4}$  & $1.39\times 10^{9}$  & 22.4  & 30 & 20      & $s$, $r$ & $\lambda_{ij}$ unmeasured for 3-6 \\
\nuc{123}{Sn} & 24.6    & $11/2^{-}$ & $3/2^{+}$  & $1.12\times 10^{7}$  & $2.4\times 10^{3}$   & 100   & 30 & 30      & $r$      & $\lambda_{ij}$ unmeasured for 3-7 \\
\nuc{125}{Sn} & 27.50   & $11/2^{-}$ & $3/2^{+}$  & $8.33\times 10^{5}$  & $5.71\times 10^{2}$  & 100   & 30 & 30      & $r$      & $\lambda_{ij}$ unmeasured for 3-8 \\
\nuc{127}{Sn} & 5.07    & $11/2^{-}$ & $3/2^{+}$  & $7.56\times 10^{3}$  & $2.48\times 10^{2}$  & 100   & 30 & 30      & $r$      & $\lambda_{ij}$ unmeasured for 3-8 \\
\nuc{128}{Sb} & 0.0+X   & $8^{-}$    & $5^{+}$    & $3.26\times 10^{4}$  & $6.25\times 10^{2}$  & 96.4  & 9  & unknown & $r$      & $E_m$ unknown; Note 2 below \\
\nuc{170}{Ho} & 120     & ($6^{-}$)  & ($1^{+}$)  & $1.66\times 10^{2}$  & $4.3\times 10^{1}$   & 100   & 2  & unknown & $r$      & Note 3 below \\
\nuc{176}{Lu} & 122.845 & $7^{-}$    & $1^{-}$    & $1.19\times 10^{18}$ & $1.32\times 10^{4}$  & 100   & 30 & 10      & $s$      & $\lambda_{ij}$ unmeasured for 5-13, 16, 17 \\
\nuc{182}{Hf} & 1172.87 & $0^{+}$    & ($8^{-}$)  & $2.81\times 10^{14}$ & $3.69\times 10^{3}$  & 54    & 30 & 10      & $s$, $r$ & $\lambda_{ij}$ unmeasured for 2-9, 11, 12 \\
\enddata

\tablenotetext{1}{Astrophysical site key: $p$: $p$-process.  $r$: $r$-process.  $s$: $s$-process.  Sne: supernovae.  PSne: pre-supernovae.}
\tablenotetext{2}{Intermediate states unmeasured.}
\tablenotetext{3}{$E_m$ uncertain; intermediate states of \nuc{170}{Ho} unmeasured and \nuc{170}{Dy} $\beta$ intensities unmeasured.}

\label{tab:data_summary}

\end{deluxetable*}

We provide ASCII formatted data files in a .tar.gz package containing the effective transition and weak-interaction rates for the isotopes presented in this paper. The rates are compiled into two data files for each isotope (one for the ground state and one for the isomer) along with relevant metadata, including publication data, the nuclear level energy, the assumed density of the environment, and descriptions of the columns.  We always take $\rho Y_e = 10^5$ g/cm$^3$, which should be adequate for most environments where electrons are not degenerate and electron capture is not the dominant weak interaction.  We also provide the rates from our sensitivity studies of {\Al} and {\Kr}.  This data is intended for unlimited release under Los Alamos report LA-UR-20-26010.

\section{Acknowledgments}

We thank Frank Timmes, Aaron Couture, Brad Meyer, Alex Heger, and George Fuller for helpful astrophysics discussions.  Calvin Johnson deserves credit for the term ``astromer''.  Timothy Elling, Jonah Miller, Peter Polko, Joseph Schaeffer, and Marc Verriere provided insights on pathfinding.  Alice Shih helped create figure \ref{fig:digraph_cartoon}.  G.W.M. would like to acknowledge Presh Talwalkar and his YouTube channel Mind Your Decisions for inspiring equations \ref{eq:prob_comp} and \ref{eq:prob_vec}. 
G.W.M and M.R.M. were supported by the US Department of Energy through the Los Alamos National Laboratory. Los Alamos National Laboratory is operated by Triad National Security, LLC, for the National Nuclear Security Administration of U.S.\ Department of Energy (Contract No.\ 89233218CNA000001). 
G.W.M and M.R.M. were also supported by the Laboratory Directed Research and Development program of Los Alamos National Laboratory under project number 20190021DR. 
This research was partially supported by the National Natural Science Foundation of China (No. U1932206) and the National Key Program for S\&T Research and Development (No. 2016YFA0400501).

\software{WebPlotDigitizer \citep[v4.3;][]{Rohatgi2020},
MatPlotLib \citep[v3.3.1; ][]{Hunter07},
NumPy \citep[v1.19.0; ][]{2011CSE....13b..22V},
SciPy \citep[v1.5.2; ][]{2020SciPy-NMeth}
}

\appendix

\section{Transition Pathways}
\label{app:pathfinding}

In determining the effect of unmeasured individual transition rates on total effective transition rates, it is helpful to identify those paths through intermediate states which a nucleus is most likely to follow.  We describe here the probabilities to follow specific paths, the algorithm we employed to identify the dominant paths, and comment on the symmetry of reverse paths.

\subsection{Path Probabilities}
\label{sec:path_prob}

The probability to follow a path is the product of the weights $b_{ij}$ for each step in the path.  We will show here precisely what that means.  A probability quantifies how frequently an outcome will occur from a set of possibilities, and the probabilities must sum to unity.  In discussing path probabilities, we must define the possibilities under consideration.

The first set of possibilities we discuss is not directly applicable to calculations of effective transition rates, but rather is useful in interpreting the $P_{iE}$.  Consider the set of all paths of length $N$ which begin at state $i$ and end anywhere.  Naturally, the probabilities to follow these paths must sum to unity, since the system must end up \emph{somewhere}.  For paths of length 1, we can easily see that the probability to follow each path to each final state $f$ is $b_{if}$.  Those are the individual transition probabilities, and we know that they sum over $f$ to unity.  Now consider paths of length 2: $i \rightarrow j \rightarrow f$.  We know the probabilities of the individual transitions, so we may simply multiply them, then sum the products over all values of $j$ and $f$ to get a normalization factor $K$ for the total probability.
\begin{align}
    K = \sum\limits_j b_{ij} \sum\limits_f b_{jf}
\end{align}
The inner sum is unity, which implies the the total sum is unity, and the normalization factor is simply 1.  Considering paths of greater length simply extents the number of sums, and we see that paths of length $N$ form a well-defined set of possibilities with probabilities computed from direct multiplication of the individual transition probabilities.

Our graph represents an irreducible Markov chain (every state is reachable from every other state by at least one path), so all states are members of a single communicating class \citep{gagniuc2017markov}.  Consequently, each intermediate state can reach $A$ via at least one path with a finite number of steps $N$, and that path has positive probability with respect to other paths of length $N$.  This implies that with repeated transitions the probability to \emph{not} have followed a path that reaches $A$ will decay toward zero.  Thus $A$ is recurrent, meaning the system will eventually reach $A$ with unit probability.  Recurrence is a class property (all members of a communicating class are either recurrent or not recurrent), so a system will eventually reach $A$ with probability 1 \emph{and} it will eventually reach every other endpoint state with probability 1.

The $P_{iE}$ from section \ref{sec:rate_formalism} quantify the probability of starting from intermediate state $i$ and reaching endpoint $E$ before reaching any other endpoint.  The system definitely reaches every endpoint eventually, and it must reach one of them first; the probabilities $P_{iE}$ therefore sum to 1.
\begin{equation}
    \sum\limits_E P_{iE} = 1
    \label{eq:P_sum}
\end{equation}

Now we describe a more directly applicable set of possibilities.  We are interested in starting at $A$, leaving to some state $s$ (which in principle could be $B$), and finding the subsequent probability to successfully transition to endpoint state $B$.  Again, for every state $s$, the sum over $t$ of $b_{st}$ is unity, which combines with equation \ref{eq:P_sum} to give a set of possibilities whose probabilities sum to unity.
\begin{equation}
    \sum\limits_s \left( b_{As} \sum\limits_E P_{sE} \right) = \sum\limits_s b_{As} = 1
    \label{eq:prob_sum}
\end{equation}

In words, the set of possibilities for which we compute probabilities is paths that begin at $A$, terminate at an endpoint state, and do not have any endpoint states in between.  When the system leaves $A$, it must eventually reach an endpoint state (equation \ref{eq:P_sum}), so the probabilities of all possibilities sum to 1 (equation \ref{eq:prob_sum}).  The possibilities are also mutually exclusive (no two paths are alike), giving a well-defined set of possibilities and associated probabilities.

Iterated expansion of the $P_{jB}$ in equation \ref{eq:P_recursion} reveals that $P_{iB}$ is the sum over all paths of any length from $i$ to $B$ of the compounded individual transition probabilities.
\begin{equation}
    P_{iB} = \sum\limits_{\substack{{\rm paths} \\i \rightarrow j \rightarrow ... \rightarrow k \rightarrow B}} b_{ij}...b_{kB}
\end{equation}
Note that this includes short paths which might not have a $j$ or $k$.  Multiplying equation \ref{eq:P_recursion} by $b_{Ai}$ gives an un-normalized probability that when the system leaves $A$, it goes to $i$, and from there via some path that eventually leads to $B$, with the contribution of each path to that probability explicitly computable.  Likewise, we may compute the un-normalized probabilities of paths that begin at $A$ and go to any endpoint $E$ (including returning to $A$) by using the $P_{iE}$.  But comparison with equation \ref{eq:prob_sum} shows that the sum of the compounded individual transition probabilities for all paths that start at $A$ and end at any endpoint is unity.  In other words, the compounded individual transition probabilities for these paths directly answer the question ``When the system leaves $A$, what is the probability to follow a particular path to a particular endpoint state given that the path terminates at an endpoint state?''  Naturally, we may replace $A$ with any endpoint state.

The total effective transition rate $\Lambda_{AB}$ is ultimately the product of the rate $\lambda_A$ at which systems leave $A$ and the sum of the probabilities of paths from $A$ to $B$ (see equation \ref{eq:Lam_eff} in the context of the arguments in this section).  Each path $A \rightarrow ... \rightarrow B$ contributes to the total effective transition rate in proportion to its probability, and we may now ask which paths contribute the most to transition rates.

\subsection{Pathfinding Algorithm}
To find the most probable paths through intermediate states, we use a version of the A$^*$ pathfinding algorithm \citep{hart1968formal}.  In its original form, A$^*$ finds the \emph{single} path of least cost between two vertices that does \emph{not revisit} intermediate vertices on a weighted graph.  The edge costs are also typically \emph{non-negative and additive}.  For our purposes, each of the emphasized points requires generalizing.

The first generalization is well-known: finding the $k$ shortest paths, rather than the single shortest \citep{eppstein1998finding}.  Although $k$ is the standard symbol for finding multiple shortest paths, we will for the remainder of this section use $N$ to avoid confusion with a state labeled $k$.

The second generalization requires a notion of ``legality''.  Consider a graph which represents a geographic map, and we are looking for routes from $A$ to $B$.  We would not be interested in paths that revisit vertices since it would effectively be backtracking.  We would therefore declare any paths that include revisiting to be illegal and not consider such paths as candidates.  In general, at each iteration of the algorithm described below, we would determine whether each next increment in the path is legal according to our specific needs, then as appropriate either add it to or exclude it from the list of candidate paths.

In the class of cases under consideration here, a physical system may meander between intermediate states---even tracing loops in the graph---before reaching $B$.  If we exclude paths that revisit intermediate states, we will miss the contribution of these meanderings to the total effective transition rate.  Indeed, some paths with loops may contribute more than some direct paths.  This leads to our statement of path legality: we allow any path that starts at $A$, ends at $B$, and does not have $A$ or $B$ as intermediates.

The final generalization addresses how we compute cost.  In the graphs considered here, the edge weights are multiplicative probabilities.  These may be converted to non-negative additive weights by taking the negative logarithm.
\begin{align}
    w_{ij} &= -\log (b_{ij}) \nonumber \\
    \rightarrow w_{ij} + w_{jk} + ... &= -\log (b_{ij} \cdot b_{jk} \cdot ... )
\end{align}

But this is an unnecessary step if we change the optimization criterion from ``least cost'' to ``highest probability''.  We may just as easily compute the cumulative effect of each incremental addition to a path; we simply multiply instead of add, and we prefer paths with higher cost (probability) instead of lower.

A$^*$ requires a heuristic function $h(v)$ that estimates the cost from each vertex $v$ to the goal (here, the goal is $B$).  Candidate paths are then selected according to the criterion of optimizing the computed actual cost to get to the current vertex combined with the estimated cost to get the rest of the way to the goal.  This heuristic must be ``admissible'', meaning it never overestimates the actual cost.  With our multiplicative weights, our choice of heuristic must therefore never \emph{underestimate} the probability.  For simplicity, we take $h(v)=1 ~\forall ~v$.  In principle, we could achieve better performance with a more intelligent choice of $h(v)$, but we won't worry about that here.

Finally, we describe our algorithm in detail.  To find the $N$ most probable paths from $A$ to $B$, follow these steps.

\begin{enumerate}
\item Start an empty list of candidate paths.

\item Add to this list all possible paths of length 1 (one increment) that start at $A$.  Using the graph in figure \ref{fig:digraph_cartoon} as an example, the list will now consist of the paths $A \rightarrow i$, $A \rightarrow j$, and $A \rightarrow k$.

\item \label{item:select} Select the highest probability path from the candidate list and delete it from the list.

\item \label{item:append} If the selected path does not end at $B$, add to the candidate list all legal paths that are one increment farther along and return to step \ref{item:select}.  If, for example, the most probable candidate path is $A \rightarrow i$, delete it from the list and append to the list the candidates $A \rightarrow i \rightarrow B$, $A \rightarrow i \rightarrow j$, and $A \rightarrow i \rightarrow k$.

\item If the selected path ends at $B$, record it as the $n$-th most probable path, where $n$ is the number of most probable paths found so far.  If $N$ most probable paths have been found, you're done.  Otherwise, return to step \ref{item:select}.
\end{enumerate}

It's worth mentioning that it is in principle possible to get caught in a loop of transitions where probability decays slowly with each iteration, and a path may traverse this loop many times before it fails to be selected in step \ref{item:select}.  This can be avoided if in step \ref{item:append} we augment the legality condition to say that a path may not include more than $N-1$ total loops.  Because every path increment \emph{at best} does not increase a path's probability, we may be assured that a path $P$ which visits a state $n$ times will be at least as probable as an otherwise identical path which includes a loop to visit the state $n+1$ times.  Therefore, there will be at least $n-1$ paths at least as probable as $P$, and we may safely reject all paths with more than $N-1$ loops.

\subsection{Path Symmetry}
\label{sec:symmetry}

\begin{figure}
    \centering
    \includegraphics[width=0.45\columnwidth]{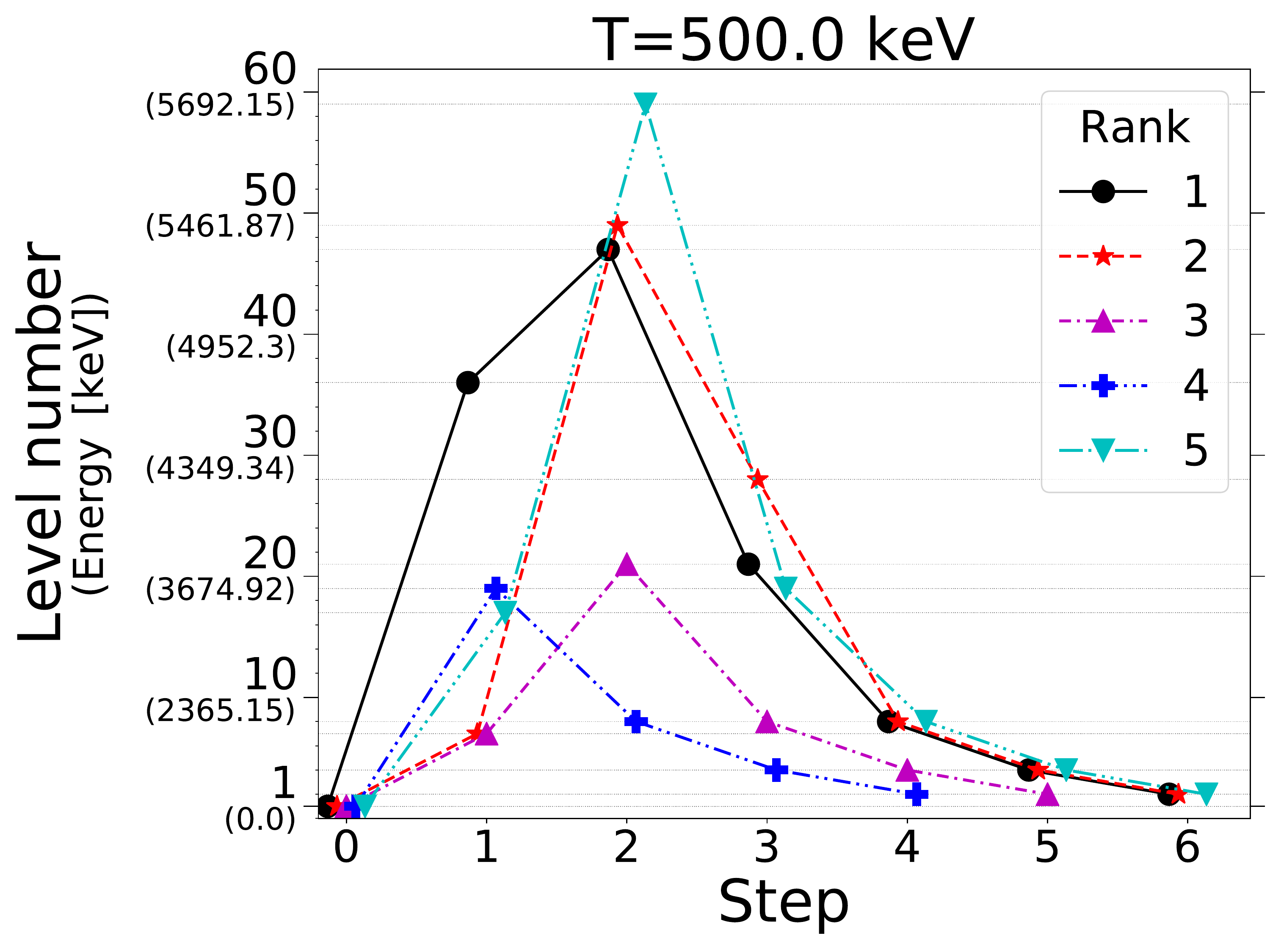}~~~~~
    \includegraphics[width=0.45\columnwidth]{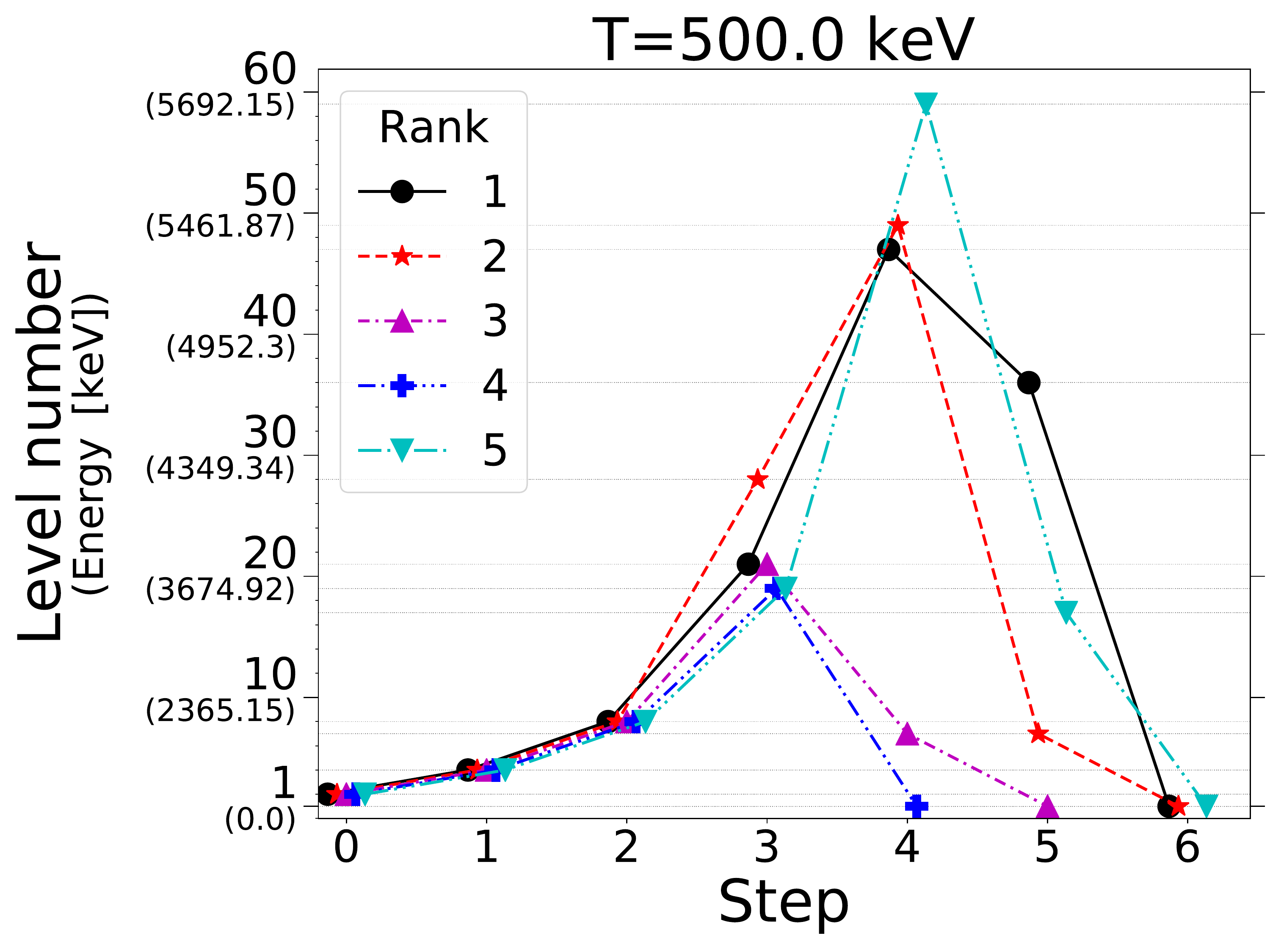}
    \caption{The five most probable paths from the ground state to the isomer (top) and from the isomer to ground (bottom) in {\Al} at a temperature $T=500$ keV.  The paths in one direction are the reverse of the paths in the other direction.}
    \label{fig:al_paths}
\end{figure}

In a general weighted digraph, the optimal path from $A$ to $B$ is not necessarily the optimal path from $B$ to $A$.  But when transitions are mediated by a thermal bath, the most probable path between two states---and hence the greatest contributor to the transition rate---is the same in both directions.  In fact, all paths maintain their relative probability when traversed in either direction.  For example, the second most probable path from $A$ to $B$ is the also the second most probable path from $B$ to $A$, and so on.  Figure \ref{fig:al_paths} illustrates this for the five most probable paths through {\Al} at temperature $T=500$ keV.  This is an unrealistically high temperature for {\Al} production, but it nicely illustrates the symmetry.  The states are labeled in increasing order of energy starting with ground = 1 and isomer = 2.  In the figure, the most probable path (rank = 1) from ground to the isomer is $1 \rightarrow 36 \rightarrow 47 \rightarrow 21 \rightarrow 8 \rightarrow 4 \rightarrow 2$, while from the isomer to ground it is $2 \rightarrow 4 \rightarrow 8 \rightarrow 21 \rightarrow 47 \rightarrow 36 \rightarrow 1$.  All other ranks exhibit the same symmetry.  We detail the inputs to our calculations in section \ref{sec:calc_al}.

A stochastic process for which all paths obey the symmetry described above is known as a reversible Markov chain (RMC), and there are at least three ways to show that nuclear energy levels with thermally mediated transitions comprise such.

First, and most simply, a system that meets the detailed balance condition is a RMC.  The detailed balance condition is that there exists a configuration of occupation probabilities for which the transition rates between every pair of states is the same forward and backward.  That is, if we denote the occupation probability of state $i$ as $n_i$, there must be some set of occupations such that $n_i\lambda_{ij} = n_j\lambda_{ji}$ for all $i$ and $j$ \citep{durrett1999essentials}.  The thermal equilibrium distribution $n_i \propto (2J_i+1)e^{-E_i/T}$ satisfies this equation (c.f. equation \ref{eq:lambda_det_bal}), guaranteeing reversibility.  Importantly, the system need not be in this configuration; the configuration must only exist.

Second, the specific symmetry here is directly provable, which we will do presently.  Third, by a trivial extension of the direct proof, we can show that thermally mediated transitions satisfy Kolmogorov's criterion, which states that a Markov chain is reversible if for every closed loop, the probability to follow the loop is the same in both directions \citep{kelly2011reversibility}.

To directly prove path reversal symmetry, we begin with the probability $p_{Ai...jB}$ that a transition out of state $A$ follows the path $A \rightarrow i \rightarrow ... \rightarrow j \rightarrow B$.

\begin{equation}
    p_{Ai...jB} = b_{Ai}\cdot b_{i...}\cdot ... \cdot b_{...j}\cdot b_{jB}
    \label{eq:path_prob}
\end{equation}
Rewrite the $b$s terms of $\lambda$s.
\begin{equation}
    p_{Ai...jB} = \frac{ \lambda_{Ai} }{\lambda_A}\cdot\frac{ \lambda_{i...} }{\lambda_i}\cdot ... \cdot\frac{ \lambda_{...j} }{\lambda_{...}}\cdot\frac{ \lambda_{jB} }{\lambda_j}
\end{equation}
We use equation \ref{eq:lambda_det_bal} to reverse the indices of the numerators, using $g_s \equiv 2J_s+1$ for brevity.
\begin{align}
    p_{Ai...jB} = &\left(\frac{g_i}{g_A}e^\frac{E_A-E_i}{T}\frac{ \lambda_{iA} }{\lambda_A}\right)\left( \frac{g_{...}}{g_i}e^\frac{E_i-E_{...}}{T}\frac{ \lambda_{...i} }{\lambda_i}\right)\times ...\times \left(\frac{g_j}{g_{...}}e^\frac{E_{...}-E_j}{T}\frac{ \lambda_{j...} }{\lambda_{...}}\right) \left(\frac{g_B}{g_j}e^\frac{E_j-E_B}{T}\frac{ \lambda_{Bj} }{\lambda_j}\right)
    \label{eq:lambda_reversal}
\end{align}
The exponents contain a collapsing sum, and the $g_{s\neq A,B}$ all cancel, yielding
\begin{equation}
    p_{Ai...jB} = \frac{g_B}{g_A}e^\frac{E_A-E_B}{T}\frac{ \lambda_{iA} }{\lambda_A}\cdot\frac{ \lambda_{...i} }{\lambda_i}\cdot ... \cdot\frac{ \lambda_{j...} }{\lambda_{...}}\cdot\frac{ \lambda_{Bj} }{\lambda_j}.
\end{equation}
We now multiply by $\frac{\lambda_B}{\lambda_B}$ and shift all of the numerators to the right.
\begin{align}
    p_{Ai...jB} = &\frac{g_B}{g_A}e^\frac{E_A-E_B}{T}\times\frac{1}{\lambda_A}\cdot\frac{ \lambda_{iA} }{\lambda_i}\cdot\frac{ \lambda_{...i} }{\lambda_{...}}\cdot ... \cdot\frac{ \lambda_{j...} }{\lambda_j}\cdot\frac{ \lambda_{Bj} }{\lambda_B}\cdot\lambda_B
\end{align}
Rewrite the $\lambda$s in terms of $b$s.
\begin{equation}
    p_{Ai...jB} = \frac{g_B}{g_A}e^\frac{E_A-E_B}{T}\frac{\lambda_B}{\lambda_A}b_{iA}\cdot b_{...i}\cdot ... \cdot b_{j...}\cdot b_{Bj}
\end{equation}
Finally, we recognize that this product of $b$s is the reverse path probability.
\begin{equation}
    p_{Ai...jB} = \frac{g_B}{g_A}e^\frac{E_A-E_B}{T}\frac{\lambda_B}{\lambda_A}p_{Bj...iA}
    \label{eq:path_rev_prob}
\end{equation}

The factor relating the forward and reverse path probabilities is \emph{path independent}, and depends only on temperature and the properties of the endpoint states.  Therefore, the fractional contribution of each path to transition rates from endpoint to endpoint is the same forward and backward; if a path contributes 10\% of the rate from ground to isomer, then it contributes 10\% of the rate from isomer to ground.  Reversal factors which are not equal to unity imply that one endpoint state or the other is more likely to fail to transition, returning to its starting point.

To demonstrate satisfaction of Kolmogorov's criterion, we need only set $A=B$ in the direct proof.  Since the proof does not rely on any special properties of $A$ and $B$ (e.g. longevity), $A$ and $B$ may be taken to be arbitrary and equal.  Then, per equation \ref{eq:path_rev_prob}, the reverse probability is identical to the forward probability for all closed loops.

As a final point, the rate to follow a particular path starting at endpoint $E$ is the product of the rate to leave $E$ and the probability to follow that path.  That is, $\lambda_{path} = \lambda_{E}~p_{path}$.  Note from equation \ref{eq:path_rev_prob} that the rates to follow each path in the forward and reverse directions obey the same relationship as the direct transition rates in equation \ref{eq:lambda_det_bal}.  Summing over all paths from $A$ to $B$ leads us to conclude that effective transition rates between endpoint states exhibit the thermal relationship.

\begin{equation}
    \Lambda_{AB} = \frac{g_B}{g_A}e^\frac{E_A-E_B}{T}\Lambda_{BA}
    \label{eq:effective_det_bal}
\end{equation}

This result is not surprising given an intuition of detailed balance, but it is nevertheless useful to have it made explicit and to observe that it does not require the nuclear levels to be in a thermal equilibrium distribution.  Indeed, relying on methods that do not include path reversal symmetry can lead to erroneous results.  \cite{reifarth2018treatment} made the seemingly reasonable assumption that all dominant paths consist of one up transition from a long-lived state followed by a cascade of down transitions through intermediate states; this is justified by the fact that down transitions are much faster than up transitions and therefore dominate the de-excitation of intermediate states.  However, from the present analysis and figure \ref{fig:al_paths_25_35}, we see that at temperature $T=35$ keV this misses the single greatest contributing path ($1 \rightarrow 3 \rightarrow 4 \rightarrow 2$) from the ground state to the isomer in {\Al}; the consequence is a drastic underestimate of the effective transition rate in an important temperature range.

\bibliography{references}

\end{document}